# Data-Driven Multiscale Topology Optimization of Soft Functionally Graded Materials with Large Deformations

Shiguang Deng [a], Horacio D. Espinosa [b], Wei Chen [b] [*1]

[a] Department of Mechanical Engineering, University of Kansas, Lawrence, KS, USA

[b] Department of Mechanical Engineering, Northwestern University, Evanston, IL, USA

**Abstract**

Functionally Graded Materials (FGMs) made of soft constituents have emerged as promising material-structure systems in potential applications across many engineering disciplines, such as soft robots, actuators, energy harvesting, and tissue engineering. Designing such systems remains challenging due to their multiscale architectures, multiple material phases, and inherent material and geometric nonlinearities. The focus of this paper is to propose a general topology optimization framework that automates the design innovation of multiscale soft FGMs exhibiting nonlinear material behaviors under large deformations.

Our proposed topology optimization framework integrates several key innovations: (*i*) a novel microstructure reconstruction algorithm that generates composite architecture materials from a reduced design space using physically interpretable parameters; (*ii*) a new material homogenization approach that estimates effective properties by combining the stored energy functions of multiple soft constituents; (*iii*) a neural network-based topology optimization that incorporates data-driven material surrogates to enable bottom-up, simultaneous optimization of material and structure; and (*iv*) a generic nonlinear sensitivity analysis technique that computes design sensitivities numerically without requiring explicit gradient derivation. To enhance the convergence of the nonlinear equilibrium equations amid topology optimization, we introduce an energy interpolation scheme and employ a Newton-Raphson solver with adaptive step sizes and convergence criteria. Numerical experiments show that the proposed framework produces distinct topological designs, different from those obtained under linear elasticity, with spatially varying microstructures. Our framework's effectiveness and robustness are validated across a range of objective functions and boundary conditions involving large deformations.

## 1. Introduction

Soft materials have shown significant potential in various applications, such as soft robots [1,2], compliance mechanisms [3], soft actuators [4-6], and tissue engineering [7,8]. However, their integration in FGMs [9-12], advanced composite material systems with spatially varying composition and microstructural architecture designed to achieve tailored functional properties, remains limited. Such limitation is primarily due to the reliance on engineering intuition and trial-and-error methods in designing soft FGMs, which are often inefficient and error-prone. A more efficient and systematic approach is offered by topology optimization (TO), a computational design method originating from structural optimization. TO aims to determine the optimal material layout within a given design space to achieve desired mechanical or multifunctional performance while satisfying specified constraints [13]. Over the past three decades, TO has progressed from an academic concept to a widely adopted industrial design tool, contributing to innovation in diverse areas such as aerospace [14], automobile [15], energy storage [16], heat transfer [17], and battery [18].

Three primary TO methods have been developed in literature: Solid Isotropic Material with Penalization (SIMP) [13], the Level-Set Method (LSM) [19,20], and the Evolutionary Structural Optimization (ESO) [21]. While differing in implementation, these approaches share similar formulations: they define material layouts using Heaviside functions that map a scalar field over the design domain to values between zero (void) and one (solid) [22].

Early developments in TO often assumed linear elastic behavior to simplify analysis. While linear elasticity can offer reasonable approximations for small deformations, it fails to accurately capture the mechanical response of materials subjected to large or extreme deformations, such as those encountered in compliant mechanisms [3], finite strain elasticity [23], plasticity [24], and damage [25]. Using linear models in these contexts can lead to misleading or unfaithful designs. To address this, it is essential to account for both material and geometric nonlinearities. Early efforts in this direction [13,26,27] explored the distinctions between linear and nonlinear TOs in terms of problem formulation and sensitivity analysis. More recently, there has been growing interest in tackling nonlinear TO using SIMP-based methods [28], LSM [29], and ESO techniques [30,31].

[1] *Corresponding author.
Email address: weichen@northwestern.edu

Numerous studies have shown that TOs involving large deformations can lead to severely distorted or even inverted mesh elements, particularly in regions with low stiffness. These distortions often cause numerical instability by compromising the positive definiteness of the nonlinear equilibrium equations during iterative solvers, such as the Newton-Raphson method [32]. Addressing this low stiffness-related instability is therefore a critical issue in nonlinear TOs. Several strategies have been proposed to mitigate these challenges. One approach excludes the calculation of internal forces at nodes surrounded by low-stiffness elements [33], while another research showed the effectiveness of removing and reintroducing low-density elements to aid convergence [34]. Another approach used an adaptive convergence threshold to improve convergence performance [28]. Researchers in [35] demonstrated the feasibility of introducing element connectivity parameters to connect elements via fictitious springs while eliminating low density elements. Another technique scaled local displacements within a controlled range to prevent excessive element deformation during optimization [36]. Additionally, the study in [23] proposed an energy interpolation scheme that interpolated linear and nonlinear strain energy density functions to improve numerical stability in low-stiffness regions.

While significant advances have been made in nonlinear TOs, most existing studies have focused on single-scale, structure-level optimization using a single type of soft material. In contrast, limited attention has been given to nonlinear TO approaches that simultaneously design both the microscale material architecture and the macroscale structural topology using multiple soft materials. To address this research gap, we present a general multiscale, nonlinear TO framework for designing hyperelastic FGMs composed of spatially varying soft constituents undergoing large deformations. The key contributions of this work are outlined below.

- First, we develop a homogenization scheme based on stored energy functions enabling close approximation of the effective properties of microscale composites made from multiple hyperelastic constituents.
- Second, we introduce a new microstructure reconstruction algorithm that generates microscale morphologies using a low-dimensional set of design parameters.
- Third, we develop a data-driven surrogate model for hyperelastic composites. By integrating with structure TO, our framework allows for simultaneous design of material morphology and structure topology across different scales.
- Finally, we innovatively combine neural networks (NNs)-based TO with adjoint method, implicit

function theorem, and automatic differentiation in differential programming to re-parameterize nonlinear TO as an implicit function of NNs' weights, enabling automated sensitivity analysis for various design objectives and constraints.

The rest of our paper is organized as follows. In Section 2, we review the basics of nonlinear finite element procedures and hyperelastic material models, and then we discuss our homogenization scheme for microstructures containing multiple hyperelastic constituents in Section 3. We proceed to introduce our data-driven surrogate of constitutive models in Section 4, where we provide detailed descriptions on stochastic microstructure reconstruction, material dataset generation, and surrogate modeling via Gaussian Process regression. In Section 5, we present the neural networks-based TO model with a generic nonlinear sensitivity analysis procedure. In Section 6, we discuss two numerical stabilization techniques, including an elastic energy interpolation scheme, and a Newton-Raphson model with adaptive step sizes and convergence criteria, to improve the convergence behaviors of equilibrium iterations for our multiscale TOs under large deformations. In Section 7, we evaluate the performance of our framework by several numerical experiments, and we conclude the paper in Section 8.

## 2. Nonlinear Finite Element Analysis

In this section, we briefly review Finite Element Analysis (FEA) procedures for nonlinear problems involving material nonlinearity and geometric nonlinearity. In the total Lagrangian FEA framework [37], a material point $\boldsymbol{X}$ in the reference configuration is related to its current position $\boldsymbol{x}$ in the deformed configuration as:

$$\boldsymbol{x} = \boldsymbol{X} + \boldsymbol{u}(\boldsymbol{X}, t),\qquad(1)$$

where $\boldsymbol{u}$ represents the displacement vector and $t$ refers to the time. Based on the displacement $\boldsymbol{u}$, the deformation gradient tensor $\boldsymbol{F}$ can be defined as:

$$\boldsymbol{F} = \frac{\partial \boldsymbol{x}}{\partial \boldsymbol{X}} = \boldsymbol{I} + \frac{\partial \boldsymbol{u}}{\partial \boldsymbol{X}},\qquad(2)$$

where $\boldsymbol{I}$ is the second-order identity matrix. While the deformation gradient $\boldsymbol{F}$ captures the deformation of an infinitesimal material point from its reference configuration to the deformed configuration, the volume change of the deformed body can be computed by the determinant as $J = \det(\boldsymbol{F})$. Using the deformation gradient $\boldsymbol{F}$, we define a nonlinear strain field as:

$$\boldsymbol{E} = \frac{1}{2}(\boldsymbol{F}^T \boldsymbol{F} - \boldsymbol{I}),\qquad(3)$$

where $\boldsymbol{E}$ refers to the Green-Lagrangian strain tensor. Its conjugate stress can be defined by the stored energy density function $\Psi$ as:

$$\boldsymbol{S} = \frac{\partial \psi(\boldsymbol{u})}{\partial \boldsymbol{E}},\qquad(4)$$

where $\boldsymbol{S}$ represents the second Piola-Kirchhoff (PK2) stress tensor. The PK2 stress can be converted from the Cauchy stress tensor $\sigma$ as:

$$\boldsymbol{S} = J\boldsymbol{F}^{-1}\sigma\boldsymbol{F} . \qquad (5)$$

For a nonlinear FEA, its equilibrium equation can be stated as:

$$\boldsymbol{r} = \boldsymbol{f}^{ext} - \boldsymbol{f}^{int} = \boldsymbol{0} , \qquad (6)$$

where $\boldsymbol{f}^{ext}$ is the external load vector, which we assume to be independent of design variables or displacement vectors, and $\boldsymbol{f}^{int}$ is the internal load vector. We adopt the Newton-Raphson method [32] to iteratively solve the equilibrium equation with the incremental displacement solution $\Delta\boldsymbol{u}$:

$$\boldsymbol{K}\Delta\boldsymbol{u} = \boldsymbol{r} , \qquad (7)$$

where the displacement solution is updated via $\boldsymbol{u} = \boldsymbol{u} + \Delta\boldsymbol{u}$, and $\boldsymbol{K}$ is the tangent stiffness matrix, which is the sum of two stiffness matrices:

$$\boldsymbol{K} = \frac{\partial \boldsymbol{r}}{\partial \boldsymbol{u}} = \boldsymbol{K}^{mat} + \boldsymbol{K}^{geo} , \qquad (8)$$

where $\boldsymbol{K}^{mat}$ is the material tangent stiffness matrix that describes the nonlinear material responses, and $\boldsymbol{K}^{geo}$ is the geometric stiffness matrix that accounts for the geometric effects during large deformations, including rotation and stretching. Readers are referred to [37] for the detailed derivations of $\boldsymbol{K}^{mat}$ and $\boldsymbol{K}^{geo}$.

Compressible Neo Hookean Model

In this work, we assume all soft materials follow the compressible Neo Hookean constitutive model which can be considered as an extension of the isotropic linear elastic model to large deformation regimes. The stored energy density function $\Psi$ of compressible Neo Hookean elastomers can be written as:

$$\psi = \frac{1}{2}\lambda_0 \left(\ln J\right)^2 - \mu_0 \ln J + \frac{1}{2}\mu_0 \left(I_1 - 3\right) , \qquad (9)$$

where $I_1$ is the first strain invariant as $I_1 = \mathrm{trace}(\boldsymbol{C})$ and $\boldsymbol{C}$ is the right Green-Cauchy deformation tensor $\boldsymbol{C} = \boldsymbol{F}^T\boldsymbol{F}$. $\mu_0$ and $\lambda_0$ are the initial Lame parameters determined from linearized theory. In the total Lagrangian framework, we assume the Neo Hookean elastomers are isotropic with respect to their initial unstressed configurations.

Using the Equation (4), the PK2 stress tensor can be computed as:

$$\boldsymbol{S} = \lambda_0 \left(\ln J\right)\boldsymbol{C}^{-1} + \mu_0 \left(\boldsymbol{I} - \boldsymbol{C}^{-1}\right) . \qquad (10)$$

The tangent modulus $\boldsymbol{C}^{SE}$, also known as the second elastic tensor that relates the Green-Lagrangian strain tensor $\boldsymbol{E}$ with the PK2 stress $\boldsymbol{S}$, can be computed as the Hessian of the stored energy density function with respect to deformations as:

$$\boldsymbol{C}^{SE} = \frac{\partial^2 \psi}{\partial \boldsymbol{E}\partial \boldsymbol{E}} = 4\frac{\partial^2 \psi}{\partial \boldsymbol{C}\partial \boldsymbol{C}} , \qquad (11)$$

where the component form of $\boldsymbol{C}^{SE}$ can be expressed as:

$$C_{ijkl}^{SE} = \lambda C_{ij}^{-1}C_{kl}^{-1} + \mu\left(C_{ik}^{-1}C_{jl}^{-1} + C_{il}^{-1}C_{kj}^{-1}\right) , \qquad (12)$$

where $\lambda = \lambda_0$ and $\mu = \mu_0 - \lambda \ln J$. It can be seen from Equation (12) that the tangent modulus of Neo Hookean elastomers has the same form as Hook's model for small-strain elasticity, except that its shear modulus is dependent on the deformation $J$.

3. Homogenization of Hyperelastic Composite Microstructure

The previous section reviewed the Neo-Hookean model for a single material point under finite deformation. In a multiscale setting, however, each macroscale material point is typically associated with an underlying microstructure characterized by specific material compositions and morphologies. In this context, we assume that each microstructure is composed of multiple Neo-Hookean elastomers, each defined by distinct sets of Lame parameters (please see Section 6 for the values of Lame parameters). The total stored energy density function $\Psi_m$ of the microstructure is written as:

$$\psi_m = \sum_{i=1}^{n}\left[c_i\left(\rho_i, \gamma_i\right)\psi_i\left(\mu_{0i}, \lambda_{0i}, \boldsymbol{C}\right)\right] , \qquad (13)$$

where $n$ is the number of the Neo Hookean constituents in the microstructure, and $\Psi_i$ is the individual energy density function of each constituent. $\Psi_i$ depends on the overall deformation $\boldsymbol{C}$ and the Lame parameters of $\mu_{0i}$ and $\lambda_{0i}$. Additionally, $c_i$ is the coefficient for each elastomer, and we assume it is a function of volume fraction $\rho_i$ and morphology $\gamma_i$ for each constituent. We can rewrite the total stored energy density function of Equation (13) as:

$$\psi_m = \sum_{i=1}^{n}\left[c_i\left(\rho_i, \gamma_i\right)\begin{pmatrix}\frac{1}{2}\lambda_{0i}\left(\ln J\right)^2 - \mu_{0i}\ln J \\ + \frac{1}{2}\mu_{0i}\left(I_1 - 3\right)\end{pmatrix}\right] \qquad (14)$$

$$= \frac{1}{2}\lambda_m\left(\ln J\right)^2 - \mu_m \ln J + \frac{1}{2}\mu_m\left(I_1 - 3\right) ,$$

where the interpolated Lame parameters of the composite microstructure are $\mu_m = \sum_{i=1}^{n}(c_i\mu_{0i})$ and $\lambda_m = \sum_{i=1}^{n}(c_i\lambda_{0i})$. We note that in this derivation, we assume constituents are perfectly bonded. Additionally, there is no interfacial energy or coupling effects in the microstructure, and the constituents' properties are sufficiently close such that the deformations within each constituent can be approximated by the overall deformation at the associated macroscale material point. From Equation (14), we can see that the total stored energy density function of a microstructure follows the same compressible Neo Hookean model, except that its Lame parameters are now functions of the volume fraction, morphology, and initial Lame parameters of each constituent.

We proceed to use the computational homogenization scheme [38] to approximate the homogenized linearized

Lame parameters $\mu_m$ and $\lambda_m$. Specifically, the linear elastic stiffness tensor $C_{ijkl}^H$ of the microstructure under periodic boundary conditions can be written as:

$$C_{ijkl}^H = \frac{1}{|\boldsymbol{X}|} \sum_{e=1}^{N} \left(\boldsymbol{u}_e^{A(ij)}\right)^T \boldsymbol{k}_e \boldsymbol{u}_e^{A(kl)} \,, \qquad (15)$$

where $\boldsymbol{u}_e^{A(kl)}$ and $\boldsymbol{u}_e^{A(ij)}$ are the displacement vectors corresponding to unit test train fields and admissible displacement fields, respectively, and $\boldsymbol{k}_e$ is the element stiffness matrix. Equation (15) can be simplified by letting $11 \rightarrow 1$, $22 \rightarrow 2$, and $12 \rightarrow 3$ in 2D:

$$C_{ij}^H = \frac{1}{|\boldsymbol{X}|} \sum_{e=1}^{N} q_e^{(ij)} \,, \text{ and} \qquad (16)$$

$$q_e^{(ij)} = \left(\boldsymbol{u}_e^{A(i)}\right)^T \boldsymbol{k}_e \left(\boldsymbol{u}_e^{A(j)}\right), \qquad (17)$$

where $q_e^{(ij)}$ are the element mutual energies from the test and admissible element displacement solutions, $\boldsymbol{u}_e^{A(i)}$ and $\boldsymbol{u}_e^{A(i)}$, respectively. We can then trivially obtain the values of the homogenized Lame parameters $\mu_m$ and $\lambda_m$ from the effective elastic tensor $C_{ij}^H$ from Equation (16), and use them to model the hyperelastic behaviors of macroscale elements.

## 4. Data-Driven Surrogate

In this section, we present a data-driven surrogate model to approximate the effective elastic properties of composite microstructures as a function of low-dimensional design parameters. This approach is motivated by the high computational cost associated with performing multiscale TO through direct numerical methods such as FE², which are often prohibitively expensive.

Surrogate modeling offers an efficient alternative by replacing these costly microstructural simulations with fast, pre-trained surrogates. Prior studies [39-43] have demonstrated that deploying surrogates in multiscale analyses can accelerate computation by several orders of magnitude compared to classical FE² approaches. Our surrogate framework is built upon a combination of different methods, including microstructure reconstruction to generate an ensemble of statistically equivalent microstructures, Design of Experiment (DoE) for building a comprehensive material dataset through random sampling, and Gaussian Process regression for predictive modeling.

We present a novel microstructure reconstruction method in Appendix A where the generation of stochastic microstructures relies on low-dimensional descriptors from the Spectral Density Function [44]. Additionally, we develop an interface transition method in Appendix B to enable smooth boundary transition across microstructures to achieve gradually changing spatially varying compositions and properties on topologically optimized FGMs.

We use the DoE [45] to generate the dataset containing design parameters and effective elastic properties of reconstructed microstructures (see Appendix A and Appendix B). Specifically, we generate a data repository containing 1001 microstructure samples, where the ranges of design parameters, i.e., the volume fraction $\rho_m$, the SDF outer radius $R_{out}$, and the radius interval $\Delta R$ are as below:

$$\begin{cases} 0.3 \leq \rho_m \leq 0.7 \\ 15 \leq R_{out} \leq 25 \\ 0 \leq \Delta R \leq 25. \end{cases} \qquad (18)$$

The scattering plot of the design parameters is shown in Figure 1, where the value of $\rho_m$ can continuously vary between its lower and upper bounds, while the values of the $R_{out}$ and $\Delta R$ can only take integers within sampling intervals. This is because both $R_{out}$ and $\Delta R$ are the design parameters derived from the frequency components of the 2D SDF field. For a discrete microstructure image, e.g., with $500 \times 500$ pixels, it only has $500 \times 500$ discrete frequency bins, where each frequency bin is characterized by discrete spatial indices on the 2D SDF. For random sampling, we use the Sobol sequence [45] to generate the random samples for the $\rho_m$ and use full factorial method [46] to include all feasible integer values for $R_{out}$ and $\Delta R$.

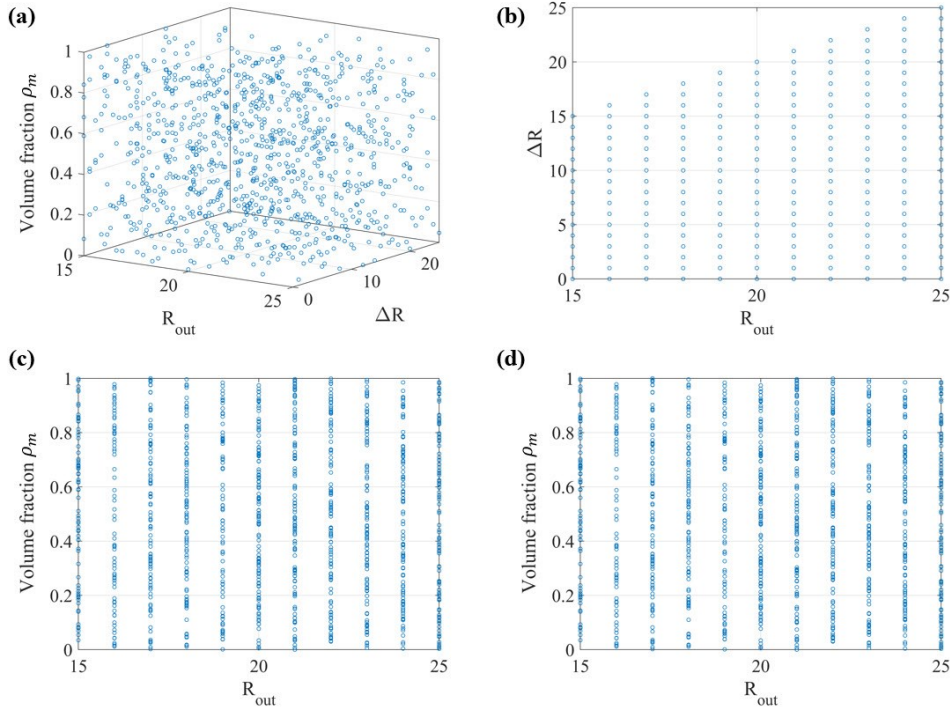

**Figure 1 DoE samples of microstructure design parameters: (a)** A total of 1001 microstructure samples are generated via DoE using the three design parameters: microstructure volume fraction ($\rho_m$), the SDF outer radius ($R_{out}$) and the SDF radius interval ($\Delta R$), where the values of $R_{out}$ and $\Delta R$ can only take integer values. The scattering plots of the random samples are projected from the 3D space onto the $R_{out} \sim \Delta R$ plane in **(b)**, the $R_{out} \sim \rho_m$ plane in **(c)**, and the $R_{out} \sim \rho_m$ plane in **(d)**, respectively.

After collecting the values of design parameters from DoE, we apply the microstructure reconstruction algorithm to rebuild morphologies and compute their effective elastic properties, i.e., the Lame parameters μ and λ, based on computational homogenization [38]. We demonstrate the distribution of the effective Lame parameters in Figure 2.

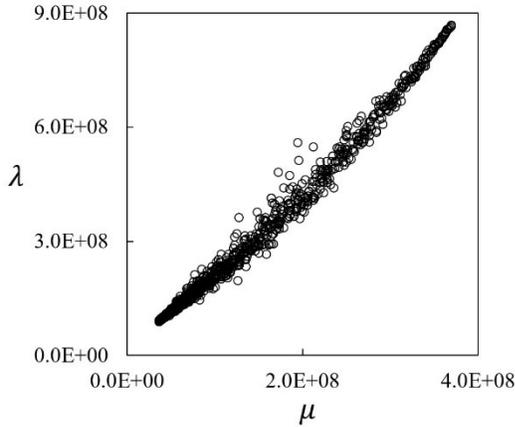

**Figure 2 Scattering plot of Lame parameters:** The homogenized values of the two Lame parameters (μ and λ) for the 1001 stochastic microstructure samples generated from DoE.

### Gaussian Process Regression

We now have a dataset of microstructure design parameters and effective Lame parameters for the 1001 samples generated from DoE. We proceed to create a Gaussian Process (GP) regression model [47-50] to emulate the relation between microstructural design parameters and their effective Lame parameters.

Per scalar quantity of interest $q$, GP considers it as a realization from the random field $Q$ over the $d_s$ dimensional input space with the input variables $\boldsymbol{s} = [s_1, s_1, ..., s_{d_s}]^T$ where the random field $Q$ is written as:

$$Q(\boldsymbol{s}) = \boldsymbol{m}(\boldsymbol{s})^T \boldsymbol{\beta} + \varepsilon(\boldsymbol{s}), \quad (19)$$

where $\boldsymbol{m}^T(\boldsymbol{s})\boldsymbol{\beta}$ is the mean as in Equation (20) and $\varepsilon(\boldsymbol{s})$ is a zero-mean GP with the covariance function defined as in Equation (21):

$$\mathrm{E}\big[Q(\boldsymbol{s})\big] = \boldsymbol{m}(\boldsymbol{s})^T \boldsymbol{\beta}, \text{ and} \quad (20)$$

$$Cov\big[Q(\boldsymbol{s}),\, Q(\boldsymbol{s}')\big] = Cov\big[\varepsilon(\boldsymbol{s}),\, \varepsilon(\boldsymbol{s}')\big] = c\big(\boldsymbol{s},\, \boldsymbol{s}'\big), \quad (21)$$

where $\boldsymbol{m}(\boldsymbol{s}) = [m_1(\boldsymbol{s}), m_2(\boldsymbol{s}), ..., m_u(\boldsymbol{s})]^T$ is a vector with $u$ known basis functions, and $\boldsymbol{\beta} = [\beta_1, \beta_2, ..., \beta_u]^T$ is a vector of $u$ unknown parameters. Since we have no prior information about the mean, we simply assume $u = 1$ and $\boldsymbol{m}(\boldsymbol{s}) = [1]^T$. In other words, our prior mean is a constant of the value of $\beta$. A kernel is applied to replace the covariance function $Cov[Q(\boldsymbol{s}), Q(\boldsymbol{s}')]$ of the random field $Q$ as the covariance of the input variables $\boldsymbol{s}$. For Gaussian covariance, the kernel $c(\boldsymbol{s}, \boldsymbol{s}')$ can be expressed as a correlation function $r(\boldsymbol{s}, \boldsymbol{s}')$, a weighted distance between $\boldsymbol{s}$ and $\boldsymbol{s}'$ as:

$$c(\boldsymbol{s}, \boldsymbol{s}') = \sigma^2 r(\boldsymbol{s}, \boldsymbol{s}') \text{, and} \tag{22}$$

$$\begin{aligned} r(\boldsymbol{s}, \boldsymbol{s}') &= exp\left[-\sum_{i=1}^{d_s} 10^{w_i}\left(\boldsymbol{s}_i - \boldsymbol{s}_i'\right)^2\right] \\ &= exp\left[\left(\boldsymbol{s} - \boldsymbol{s}'\right)^T \boldsymbol{\Omega_s}\left(\boldsymbol{s} - \boldsymbol{s}'\right)\right], \end{aligned} \tag{23}$$

where $\sigma^2$ is the prior variance of the random field $Q(\boldsymbol{s})$. $\boldsymbol{w} = [w_1, w_2, \ldots, w_{d_s}]^T$ is a vector of roughness parameters, controlling the smoothness of the random field $Q(\boldsymbol{s})$, and $\boldsymbol{\Omega_s} = diag(10^{\boldsymbol{w}})$. To estimate the model parameters $\boldsymbol{\beta}$, $\sigma^2$ and $\boldsymbol{w}$, we follow the practice of Maximum Likelihood Estimation (MLE) [47] to maximize the logarithm of the likelihood as:

$$\left[\hat{\boldsymbol{\beta}}, \ \hat{\sigma}^2, \ \hat{\boldsymbol{w}}\right] = \underset{\boldsymbol{\beta}, \sigma^2, \boldsymbol{w}}{\arg\max}\left\{\log\left[L(\boldsymbol{\beta}, \sigma^2, \boldsymbol{w}|\boldsymbol{q})\right]\right\}, \tag{24}$$

where $\hat{\bullet}$ indicates the estimated parameters, and $L$ is the likelihood of the model parameters conditioning on the vector $\boldsymbol{q} = [q(\boldsymbol{s}^{(1)}, \boldsymbol{s}^{(2)}, \ldots, \boldsymbol{s}^{(i)} \ldots, \boldsymbol{s}^{(r)})]^T$ with sampling points $\boldsymbol{s}^{(i)}$. By profiling [51], the predicted mean $\hat{q}(\boldsymbol{s}^{test})$ at an unsampled testing point $\boldsymbol{s}^{test}$ can be computed as:

$$\hat{q}(\boldsymbol{s}^{test}) = \boldsymbol{m}^T(\boldsymbol{s}^{test})\hat{\boldsymbol{\beta}} + \boldsymbol{c}^T(\boldsymbol{s}^{test})\boldsymbol{C}^{-1}\left(\boldsymbol{q} - \boldsymbol{M}\hat{\boldsymbol{\beta}}\right), \tag{25}$$

where $\boldsymbol{c}(\boldsymbol{s}^{test})$ is an $r \times 1$ vector with the $i^{th}$ element as $c(\boldsymbol{s}^{(i)}, \boldsymbol{s}^{test})$, $\boldsymbol{C}$ is a $r \times r$ matrix with the element at $(i, j)$ as $c(\boldsymbol{s}^{(i)}, \boldsymbol{s}^{(j)})$, and $\boldsymbol{M}$ is a $r \times u$ matrix with the $i^{th}$ row as $\boldsymbol{m}^T(\boldsymbol{s}^{(i)})$.

We note that the values of microstructure design parameters are normalized to the range of $[0, 1]$ to be compatible with the neural TOPO in Section 5. We proceed to test the prediction accuracy on the 201 testing samples. As shown in Figure 3, the Relative Root Mean Squared Error (RRMSE) of the GP predictions is 0.0014. Our trained GP is ready for the on-line deployment in the multiscale neural TOPO in Section 5.

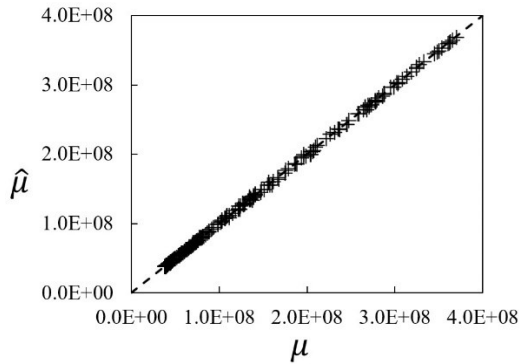

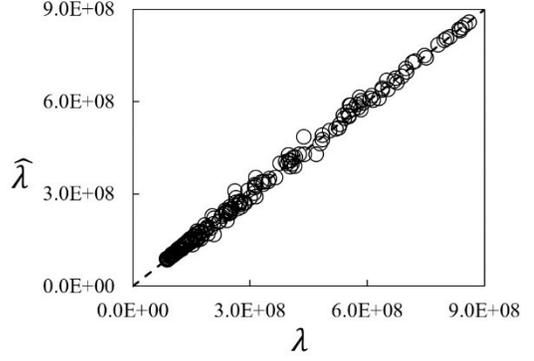

**Figure 3 GP prediction for the effective Lame parameters of microstructure samples:** The effective values of the Lame parameters (μ and λ) of 201 testing samples are predicted with an RRMSE of 0.0014.

## 5. Neural Topology Optimization

In this section, we adopt the bottom-up approach [52] to convert multiscale TO into a single-scale structure-level TO. For the single scale optimization, we use $\boldsymbol{\Theta}$ to include all design parameters as:

$$\boldsymbol{\Theta} = \left\{\rho_M, \ \rho_m, \ R_{out}, \ \Delta R\right\}. \tag{26}$$

Our TO can be formulated as:

$$\text{Minimize} \quad J = J(\boldsymbol{u}, \boldsymbol{\Theta}) \tag{27}$$

$$\text{S.T.} \quad \begin{cases} \boldsymbol{r}(\boldsymbol{u}, \boldsymbol{\Theta}) = \boldsymbol{f}^{ext} - \boldsymbol{f}^{int}(\boldsymbol{u}, \boldsymbol{\Theta}) = \boldsymbol{0} \\ g(\boldsymbol{u}, \boldsymbol{\Theta}) \leq 0, \end{cases} \tag{28}$$

where $J(\boldsymbol{u}, \boldsymbol{\Theta})$ is the design objective depending on the displacement solution $\boldsymbol{u}$ and the design parameters $\boldsymbol{\Theta}$. The design constraints in Equation (28) include an equality constraint (the nonlinear equilibrium equation) and an inequality constraint, e.g., the volume fraction constraint is defined as $g(\boldsymbol{\Theta}) = (\sum_e \rho_e A_e)/(\rho_t \sum_e A_e) - 1$. $A_e$ is the area for each macro-element, $\rho_e$ is the volume fraction of the selected constituent which is defined as $\rho_e = \rho_M \rho_m$, and $\rho_t$ is the target volume fraction. Additionally, the macro- and micro- scale volume fractions should satisfy the constraints of $0 \leq \rho_M \leq 1$ and $0 \leq \rho_m \leq 1$.

We penalize microstructural elastic properties by their macroscale volume fraction by following SIMP scheme as:

$$\tilde{\boldsymbol{C}} = \left(\boldsymbol{C}(\boldsymbol{\Theta}) - \boldsymbol{C}_{min}(\boldsymbol{\Theta})\right)\left(\rho_M\right)^p + \boldsymbol{C}_{min}(\boldsymbol{\Theta}), \tag{29}$$

where $\tilde{\boldsymbol{C}}$ represents the penalized elastic properties. And $\boldsymbol{C}_{min} = 1e^{-6}\boldsymbol{C}$ is used to avoid numerical instability. Amid optimization iterations, we follow the continuation approach [53] to gradually increase the value of the penalty parameter from $p = 1$ to $p = 3$, such that upon convergence $\rho_M$ is convergent to either zero or one, while the other design parameters ($\rho_m$, $R_{out}$, and $\Delta R$) are optimized without penalization.

## NNs-Based Design Parameterization

We extend the NNs-based TO (neural TO) proposed in [54] for our nonlinear TO. The core idea is to use NNs to map the centroid coordinates of macro-elements to the design parameters of their corresponding microstructures. In our implementation, the NNs takes two-dimensional spatial coordinates as input and outputs four design parameters. As illustrated in Figure 4, we employ a shallow network architecture consisting of a single hidden layer with 20 neurons.

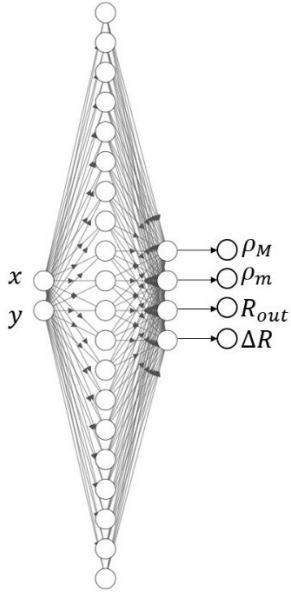

**Figure 4 NNs-based design parameterization:** Our shallow NNs (with the weight parameters $\boldsymbol{\tau}$) contain one hidden layer of 20 neurons, two input neurons corresponding to the coordinates of the centroids $(x, y)$ of macro-elements, and four output neurons accounting for the four design parameters in Equation (26).

By NNs, the design parameters $\boldsymbol{\Theta}$ can be implicitly represented as functions of the NNs' weight parameters $\boldsymbol{\tau}$ as:

$$\boldsymbol{\Theta} = \boldsymbol{\Theta}(\boldsymbol{\tau}). \tag{30}$$

We can therefore convert the TO in Equations (27) and (28) to the following formulation:

$$\text{Minimize} \quad J = J(\boldsymbol{u}, \boldsymbol{\tau}) \tag{31}$$

$$\text{S.T.} \quad \begin{cases} \boldsymbol{r}(\boldsymbol{u}, \boldsymbol{\tau}) = \boldsymbol{f}^{ext} - \boldsymbol{f}^{int}(\boldsymbol{u}, \boldsymbol{\tau}) = \boldsymbol{0} \\ g(\boldsymbol{u}, \boldsymbol{\tau}) \leq 0, \end{cases} \tag{32}$$

where both the design objective and constraints are now functions of the weight parameters of the NNs. Moreover, we apply sigmoid functions to map the networks' outputs to the range of $[0, 1]$, such that the constraints of $0 \leq \rho_M \leq 1$ and $0 \leq \rho_m \leq 1$ are satisfied by our NNs. To enforce the co-existence of materials in every microstructure, we use a linear projection function to further map the microscale volume fraction $\rho_m$ to the range of $[0.3, 0.7]$. Since $\rho_m$ is not used for material penalization in Equation (29), its value can vary freely throughout the TO iterations without incurring any penalization..

## Genetic Nonlinear Sensitivity Analysis

For gradient-based TOs, the derivation of sensitivities of linear elastic problems is often straightforward. However, the sensitivity analyses of nonlinear problems are often complex and even intractable. In this section, we introduce a generic nonlinear sensitivity analysis approach by integrating our neural TO in Section 5 with adjoint method, implicit function theorem, and automatic differentiation.

We reformulate the TO formulation in Equations (31)-(32) by combining objective with the equilibrium constraint:

$$\text{Minimize} \quad \hat{J}(\boldsymbol{\tau}) = J(\boldsymbol{u}, \boldsymbol{\tau}), \tag{33}$$

where $\hat{J}(\boldsymbol{\tau})$ is the composed objective, $\boldsymbol{u}$ is the converged displacement solution obtained by solving the equilibrium equation, and $\boldsymbol{\tau}$ is the NNs' weights in our neural TO.

The derivative of the composed objective function $\hat{J}(\boldsymbol{\tau})$ with respect to the weight parameters $\boldsymbol{\tau}$ can be written as:

$$\frac{\partial \hat{J}}{\partial \boldsymbol{\tau}} = \frac{\partial J}{\partial \boldsymbol{u}} \frac{\partial \boldsymbol{u}}{\partial \boldsymbol{\tau}} + \frac{\partial J}{\partial \boldsymbol{\tau}}. \tag{34}$$

We can also take the derivative of the equilibrium equation in Equation (6) as:

$$\frac{\partial \boldsymbol{r}}{\partial \boldsymbol{\tau}} = \frac{\partial \boldsymbol{r}}{\partial \boldsymbol{u}} \frac{\partial \boldsymbol{u}}{\partial \boldsymbol{\tau}} + \frac{\partial \boldsymbol{r}}{\partial \boldsymbol{\tau}} = 0 \text{, and} \tag{35}$$

$$\frac{\partial \boldsymbol{u}}{\partial \boldsymbol{\tau}} = -\left(\frac{\partial \boldsymbol{r}}{\partial \boldsymbol{u}}\right)^{-1} \frac{\partial \boldsymbol{r}}{\partial \boldsymbol{\tau}}. \tag{36}$$

By substituting Equations (35)-(36) into Equation (34), we obtain the sensitivity of the composed objective function as:

$$\frac{\partial \hat{J}}{\partial \boldsymbol{\tau}} = -\frac{\partial J}{\partial \boldsymbol{u}}\left(\frac{\partial \boldsymbol{r}}{\partial \boldsymbol{u}}\right)^{-1} \frac{\partial \boldsymbol{r}}{\partial \boldsymbol{\tau}} + \frac{\partial J}{\partial \boldsymbol{\tau}}. \tag{37}$$

In our work, since the number of weight parameters is much more than the number of objective functions, we can efficiently compute the sensitivity by using the adjoint method as:

$$\frac{\partial \hat{J}}{\partial \boldsymbol{\tau}} = -\boldsymbol{a} \frac{\partial \boldsymbol{r}}{\partial \boldsymbol{\tau}} + \frac{\partial J}{\partial \boldsymbol{\tau}}, \tag{38}$$

where $\boldsymbol{a}$ is the adjoint variable defined as $\boldsymbol{a}\left(\frac{\partial \boldsymbol{r}}{\partial \boldsymbol{u}}\right) = \frac{\partial J}{\partial \boldsymbol{u}}$.

It is noted that solving the adjoint $\boldsymbol{a}$ depends on the displacement solution $\boldsymbol{u}$, which is calculated via iterative Newton-Raphson procedures. Therefore, the computational cost of solving $\boldsymbol{a}$ mainly depends on the convergence performance of the Newton iterations that we will discuss in Section 6. Additionally, the computation of the adjoint $\boldsymbol{a}$ requires $\partial \boldsymbol{r}/\partial \mathbf{u}$ and $\partial J/\partial \mathbf{u}$. A naive approach to computing the two terms involves differentiation through all Newton iterations amid the back propagation. For nonlinear analyses with many Newton iterations, it is computationally demanding and memory intensive.

In our work, we adopt the implicit function theorem [55] to develop an efficient alternative with light memory footprints. We start with converting our nonlinear equilibrium of Equation (6) to a fixed-point model as:

$$\boldsymbol{u}_0 = \boldsymbol{f}\left(\boldsymbol{u}_0, \boldsymbol{\tau}_0\right), \tag{39}$$

where $\boldsymbol{f}$ is a continuous differentiable function defined by $\boldsymbol{r} = \boldsymbol{f}^{ext} - \boldsymbol{f}^{int} = \boldsymbol{0} = \boldsymbol{f}(\boldsymbol{u}_0, \boldsymbol{\tau}_0) - \boldsymbol{u}_0$. At equilibrium, we obtain a fixed point $\boldsymbol{u}_0$ of the function $\boldsymbol{f}$, where $\boldsymbol{u}_0$ is assumed to be parameterized by $\boldsymbol{\tau}_0$. In other words, upon convergence, if we continue to apply the update function $\boldsymbol{f}$, the displacement solution $\boldsymbol{u}_0$ is not changed anymore (as fixed). Then, there exist open sets $S_{\boldsymbol{u}_0}$ and $S_{\boldsymbol{\tau}_0}$ containing $\boldsymbol{u}_0$ and $\boldsymbol{\tau}_0$, respectively, and a unique continuous mapping function $\boldsymbol{u}^*$ exists in the vicinity of $\boldsymbol{\tau}_0$ such that:

$$\boldsymbol{r}\left(\boldsymbol{u}^*(\boldsymbol{\tau}), \ \boldsymbol{\tau}\right) = \boldsymbol{0}, \tag{40}$$

where $\boldsymbol{\tau} \in S_{\boldsymbol{\tau}_0}$. By differentiating both sides of Equation (40) with respect to $\boldsymbol{\tau}$ and evaluate at the equilibrium $\boldsymbol{u} = \boldsymbol{u}_0$ and $\boldsymbol{\tau} = \boldsymbol{\tau}_0$, we have the derivative of the fixed point solution as:

$$\left.\frac{\partial \boldsymbol{u}^*}{\partial \boldsymbol{\tau}}\right|_{\boldsymbol{\tau}_0} = -\left(\left.\frac{\partial \boldsymbol{r}}{\partial \boldsymbol{u}}\right|_{(\boldsymbol{u}_0, \boldsymbol{\tau}_0)}\right)^{-1} \left.\frac{\partial \boldsymbol{r}}{\partial \boldsymbol{\tau}}\right|_{(\boldsymbol{u}_0, \boldsymbol{\tau}_0)}. \tag{41}$$

By comparing the Equation (36) with Equation (41), we note that the gradient of the displacement with respect to the design parameters can be expressed as a function of the gradient of the equilibrium residuals at the convergent point. In other words, regardless of the method used to solve the nonlinear equilibrium system, the gradient of its solution can be efficiently computed using derivative information at the final converged state. This eliminates the need to backtrack through the Newton iterations, leading to a substantial reduction in both memory usage and computational cost.

We employ the automatic differentiation capability of JAX [56] to calculate the gradient fields in Equation (38) and Equation (41). The integration of the adjoint method with implicit function theorem and automatic differentiation enables us to fully automate the sensitivity computation for various nonlinear design objectives and constraints (see Section 8 for the discussion of objective functions).

## 6. Optimization Stabilization Schemes

Our TO in Equation (33) involves the equilibrium solutions solved from a nonlinear system of equations. Numerical instability often results in divergence amid equilibrium iterations. Specifically, intermediate designs in the optimization iterations often have low-density regions which are often subject to excessive deformations under extreme loading scenarios. Overly large deformations often result in severe mesh distortions that cause the tangent stiffness matrix in Equation (11) to lose positive definiteness, and erroneous values of deformation gradient, stress, and internal force. In this section, we discuss two stabilization schemes to improve the convergence performance of our nonlinear TO.

### Newton-Raphson with Adaptive Step Sizes

We adopt the Newton-Raphson procedure [32] to iteratively solve the system equilibrium equation. Here we rewrite the equilibrium equation of Equation (6) as:

$$\boldsymbol{r}\left(\boldsymbol{u}_n, t_n\right) = \boldsymbol{f}^{ext} - \boldsymbol{f}^{int}\left(\boldsymbol{u}_n, t_n\right) = \boldsymbol{0}, \tag{42}$$

where the convergence of the Newton algorithm pertains to the iterative solution of displacement $\boldsymbol{u}_n$ at the time step (loading step) $t_n$ to enforce the balance between internal forces $\boldsymbol{f}^{int}$ and external load $\boldsymbol{f}^{ext}$.

We adopt the full Newton algorithm that computes the tangent stiffness matrix at each Newton iteration. Compared to modified Newton methods where tangent matrices are only computed at the beginning of a loading step (or intermittently during a step), the full Newton approach is less efficient but more robust. Additionally, for Newton models with a fixed-size time step, it is nontrivial to determine the optimal value of step sizes: an overly conservative step size leads to high computational costs with a significant number of iterations, while an excessive step size results in excessively large boundary conditions causing inverted elements with negative determinants of deformation gradients.

We therefore develop an adaptive scheme to automatically adjust Newton's step size to accommodate convergence behaviors. We initialize our Newton model with an initial time point $t_0 = 0$, the final time point $t_f = 1$, the initial time increment $\Delta t_0 = 0.01$, the maximum $\Delta t_{max} = 0.05$ and the minimum thresholds $\Delta t_{min} = 1e^{-6}$ of the time increment. We adopt two convergence criteria to terminate Newton iterations, i.e., a force residual-based criterion and an incremental displacement-based criterion:

$$\|\boldsymbol{r}_n\| \leq \delta \max\left(\left\|\boldsymbol{f}^{ext}\right\|, \ \left\|\boldsymbol{f}_n^{int}\right\|\right), \quad \|\Delta \boldsymbol{u}_n\| \leq \delta \|\boldsymbol{u}_n\|, \tag{43}$$

where $\delta$ is the convergence tolerance, and the subscript $n$ indicates the time (loading) step. The choices of the norm $\|\cdot\|$ and the value of $\delta$ may have a significant influence on the convergence behaviors. We adopt the $l_2$ norm to control the mean error over all degrees of freedom. We set up a continuation scheme to gradually change $\delta$ as:

$$\delta_n = t_n \delta_f + \left(t_f - t_n\right)\delta_0, \tag{44}$$

where we assume the tolerance $\delta_n$ at the time step $t_n \in \left[t_0, t_f\right]$ is linearly interpolated between an initial value $\delta_0 = 0.1$ and its final value $\delta_f = 0.001$. Additionally, we set the incremental time step as:

$$\Delta t_n = \min\left(\Delta t_{max}, \ \Delta t_n, \ t_f - t_n\right), \tag{45}$$

where the time increment at the time step $t_n$ is adaptively increased to $\Delta t_n = 1.5\Delta t_{n-1}$ if the previous step at $t_{n-1}$ is convergent or reduced as $\Delta t_n = 0.25\Delta t_{n-1}$ if divergent.

Energy Interpolation Between Constitutive Models

In density-based TOs, low-density elements may undergo excessive distortions or even become distorted, which causes element tangent moduli to lose positive definiteness in geometrically nonlinear analysis and numerical instability amid optimization iterations. A simple and effective approach to address such numerical challenge is the energy interpolation scheme proposed in [23] that interpolates the energy density functions between the hyperelastic and linear elastic constitutive models depending on element densities. The elements with high density values are modeled by hyperelastic stored energy function, while the elements with low density values are simulated via the linear elastic energy function. Therefore, the interpolated energy density $\Psi_{int}(\boldsymbol{u}_e)$ is written as:

$$\psi_{int}(\boldsymbol{u}_e) = \psi_N(\kappa_e \boldsymbol{u}_e) - \psi_L(\kappa_e \boldsymbol{u}_e) + \psi_L(\boldsymbol{u}_e), \qquad (46)$$

where $\Psi_N$ is the Neo Hookean energy function, while $\Psi_L$ is the linear elastic energy density function. For the element with high density, its interpolation coefficient $\kappa_e$ are set as one ($\kappa_e = 1$), such that the interpolated energy $\Psi$ is reduced to the Neo Hookean energy $\Psi = \Psi_N$. For void and low-density elements, we set $\kappa_e = 0$, such that the stored energy corresponds to the linear elastic energy as $\Psi = \Psi_L$. Under this setting, nonlinear analysis only works for high-density elements, which tend not to distort excessively. Conversely, the distorted low-density elements are modeled by linear elasticity without numerical instability, as only the initial configuration is needed.

To ensure a smooth transition between the two constitutive models, the interpolation coefficient $\kappa_e$ can be expressed as a Heaviside projection function of the penalized element density $(\rho_M)^p$ as:

$$\kappa_e = \frac{\tanh\left(\beta_\kappa \rho_0\right) + \tanh\left(\beta_\kappa \left((\rho_M)^p - \rho_0\right)\right)}{\tanh\left(\beta_\kappa \rho_0\right) + \tanh\left(\beta_\kappa \left(1 - \rho_0\right)\right)}, \qquad (47)$$

where $p$ is the penalty parameter from the SIMP scheme. While $\beta_\kappa$ is a user-defined large constant, and $\rho_0$ is a threshold to determine element behaviors. For elements with $(\rho_M)^p < \rho_0$, they are modeled as linear elastic materials as $\kappa_e = 0$. When elements have the density of $(\rho_M)^p > \rho_0$, they follow the nonlinear elastic model as $\kappa_e = 1$. However, for elements that $0 < \kappa_e < 1$, their behaviors are interpolated between the linear and nonlinear constitutive models. We follow the values of $\beta_\kappa = 500$ and $\rho_0 = 0.01$, which are suggested by [23]. With the energy interpolation, the sensitivity of the equilibrium equation in Equation (6) is computed as:

$$\frac{\partial \boldsymbol{r}}{\partial \rho_M} = -\frac{\partial \boldsymbol{f}^{int}}{\partial \rho_M} - \frac{\partial \boldsymbol{f}^{int}}{\partial \kappa_e} \frac{\partial \kappa_e}{\partial \rho_M}. \qquad (48)$$

## 7. Numerical Experiments

We perform several numerical experiments in this section to validate our proposed multiscale nonlinear TOPO framework. Our experiments include a cantilever beam in Section 7.1, a double-clamped beam in Section 7.2, and a square beam in Section 7.3. In experiments, we demonstrate the distinction between linear elastic and hyperelastic TOPOs and the difference between single-scale designs and their multiscale counterparts under large deformations.

We assume the presence of two types of hyperelastic constituents in composite microstructures, as detailed in Section 4. Both constituents are modeled using the isotropic, compressible Neo-Hookean constitutive model, with distinct initial Lame parameters as:

$$\begin{aligned} \mu_0^A &= 3.70e8, & \lambda_0^A &= 8.64e8, \\ \mu_0^B &= 3.70e7, & \lambda_0^B &= 8.64e7, \end{aligned} \qquad (49)$$

where we assume the Lame parameters of the elastomer A are higher than the elastomer B. We note that all quantities of interest, such as modulus, length, force, objective and constraint values, are treated as dimensionless in this work for simplicity and generality.

In our experiments, we solve the TO by following Equations (31)-(32). Specifically, we assume that the inequality constraint in Equation (32) corresponds to the volume fraction of the elastomer A as:

$$g(\boldsymbol{\tau}) = \frac{\sum_e \rho_M(\boldsymbol{\tau})\rho_m(\boldsymbol{\tau})A_e}{\rho_t \sum_e A_e} - 1, \qquad (50)$$

where the target volume fraction $\rho_t$ is set as 0.3 for all experiments.

To demonstrate the robustness of our proposed TO model, we consider two types of objective functions:

$$J_1(\boldsymbol{\tau}) = \left(\boldsymbol{f}^{ext}\right)^T \boldsymbol{u}(\boldsymbol{\tau}), \text{ and} \qquad (51)$$

$$J_2(\boldsymbol{\tau}) = \int_{\Omega_0} \psi_{int}\left(\boldsymbol{u}(\boldsymbol{\tau}), \boldsymbol{\tau}\right)dX - \left(\boldsymbol{f}^{ext}\right)^T \boldsymbol{u}(\boldsymbol{\tau}), \qquad (52)$$

where $J_1$ refers to the classic end-compliance minimization problem [13], and $J_2$ corresponds to the total potential energy [57]. We aim to minimize the end compliance $J_1$ or maximize the energy $J_2$ in our experiments. Note we do not need to derive the analytical expressions for the gradients of both objectives, since they are automatically available through our generic nonlinear sensitivity analysis as discussed in Section 5. Additionally, in the experiments, we consider two types of boundary conditions for large deformation analyses: a Neumann boundary with an external force $F$, and a Dirichlet boundary $D$ with a prescribed nonzero displacement.

We generate our material dataset containing the design parameters of microstructures and their effective elastic properties from the DoE and microstructure reconstruction procedures as discussed in Section 4 and Appendix A by programming in MATLAB [58]. Our neural TOPO in Section 5 is implemented by JAX [56] to make use of its automatic differentiation for nonlinear sensitivity analysis and Just-In-Time (JIT) compilation for high-speed program execution. We perform all the numerical experiments on a 64-bit Linux workstation with 18 Intel Xeon W-2295 CPU cores running at 3.0 GHz with 256 GB of RAM.

### 7.1. Cantilever Beam

To demonstrate the robustness of our proposed model, we first apply it to design a cantilever beam as shown in Figure 5(a). For FEA, its design domain is discretized by an $80 \times 20$ mesh of quadrilateral finite elements. Each element is assumed to have a unit length and has a full integration of four Gaussian points. The domain is subject to a fixture boundary condition at left edge, and a vertical downward force $F$ is applied at bottom right corner. For TOPO, we set the design objective as the end compliance in Equation (51), while the constraint is set as the volume fraction as in Equation (50).

We start with applying a small magnitude of force $F = 1e5$ to validate our model by comparing the designs from a single-scale linear elastic TOPO with a hyperelastic counterpart. The two optimized structures are demonstrated in Figure 5(b) and Figure 5(c), respectively. We observe that the two designs have similar material layouts, which is expected, as the Neo-Hookean model converges to linear elastic law in the limit of small deformations.

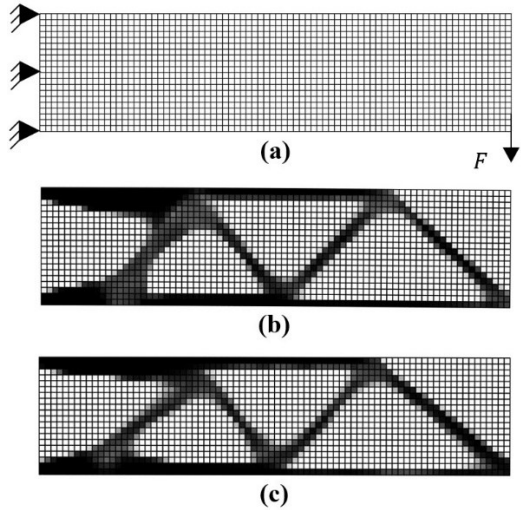

**Figure 5** Design setup of cantilever beam experiment: (a) The design domain is discretized by $80 \times 20$ quadrilateral finite elements of unit length. The beam is subject to a fixture boundary condition on the left edge and a vertical downward force ($F$) at the bottom of the right edge; (b) The deformed structural topology of a linear elastic TOPO when subject to $F = 1e5$; (c) The deformed topology after a singlescale hyperelastic TOPO with $F = 1e5$.

As we continue to increase the magnitude of the force $F$, the optimized topology tends to deviate from the linear elastic design, as shown in Figure 6. We demonstrate the optimized structures from the single-scale hyperelastic TOPO when the force is increased to $F = 1e6$ and $F = 1e7$, as shown in Figure 6(a) and Figure 6(c), respectively. It is observed that, while the designs for small deformations in Figure 5(c) and Figure 6(a) have truss-like regions of similar thickness, the design for large deformation in Figure 6(c) exhibit a different topology where the horizontal truss-like region is removed and other regions are strengthened with larger thicknesses. The horizontal truss region is excluded from the final design, as it would experience compressive forces and is prone to buckling under large deformations, a trend similarly observed in the results presented in Section 7.2.

To compare the difference between single-scale and multiscale hyperelastic TOPOs, we assume each macro-element is associated with a composite microstructure. We demonstrate the multiscale designs for the two forces $F = 1e6$ and $F = 1e7$ in the Figure 6(b) and Figure 6(d), respectively. We note that the grey color scale in the Figure 6(b) and Figure 6(d) indicates the volume fraction of the stiffer constituent A (see the moduli definition in Equation (49)) in the bi-phase composite microstructures. Compared with single-scale designs, the multiscale designs tend to allocate more stiffer materials along the top and bottom edge regions that are subject to large tension and compression forces. Compared to the design of small deformation in Figure 6(b), the large-deformation design in Figure 6(d) employs higher volume fractions of stiffer materials in the truss-like region to better resist the downward force.

For the four designs from Figure 6, we compare their elastic compliance values as in Table 1. Compared with single-scale designs, both multiscale designs have smaller elastic compliance, which is consistent Figure 6 that the vertical displacements of the multiscale designs appear smaller than the single-scale designs. This is because in the TOPO formulation in Equation (32), we set a 30% volume fraction constraint for the stiffer constituent for both single scale and multiscale designs, the usage of the softer constituent however is unconstrained in multiscale design; thus, resulting in multiscale structures with more than 30% material volume fractions, exhibiting stronger resistance to external forces, and lower elastic compliance.

**Table 1** Values of elastic compliance for the four TOPO designs in Figure 6.

|                  | $F = 1e6$ | $F = 1e7$ |
| ---------------- | --------- | --------- |
| Single-scale TOPO | 1.30e7    | 8.88e8    |
| Multiscale TOPO   | 6.64e6    | 5.52e8    |

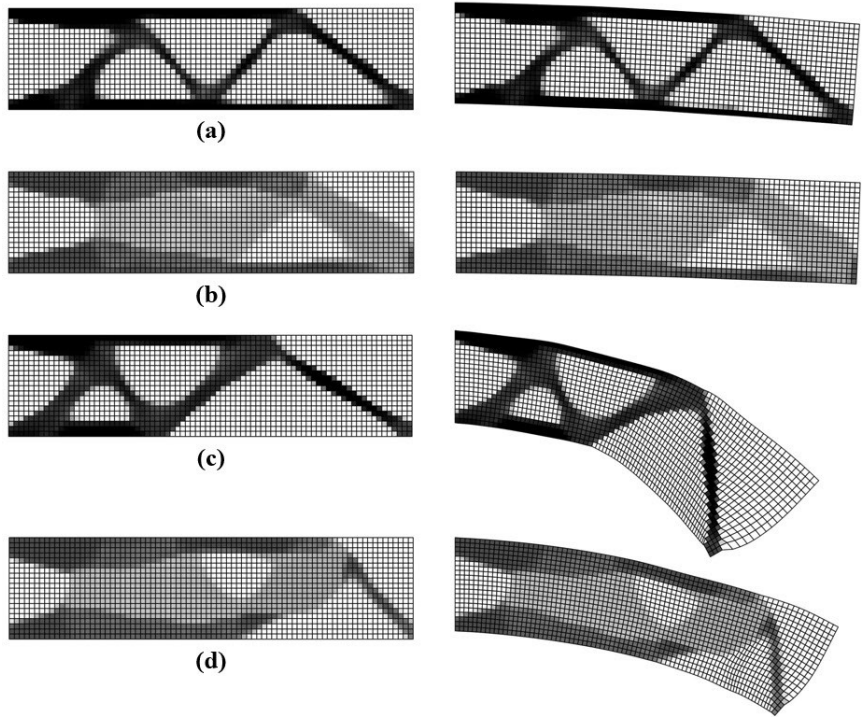

**Figure 6 Comparison of single-scale and multiscale hyperelastic TOPO when subject to different magnitudes of loading boundary conditions: (a)** The undeformed and deformed topologies of a single-scale TOPO with $F = 1e6$; **(b)** The undeformed and deformed topologies of a multiscale TOPO with $F = 1e6$ where each macro-element is assumed to be associated with a composite microstructure; **(c)** The undeformed and deformed topologies of a single-scale TOPO with $F = 1e7$; **(d)** The undeformed and deformed topologies of a multi-scale TOPO with $F = 1e7$. The grayscale indicates the volume fraction percentage of the stiffer composite constituent.

We plot the convergence histories of the elastic compliance and volume fraction as in Figure 7(a) and Figure 7(b) to help understand the optimization process for the multiscale hyperelastic TOPO in Figure 6(d). In Figure 7(a), along with the compliance history, we demonstrate four snapshots of structural topologies at different optimization iterations. We observe that the compliance initially increases with the number of iterations, then gradually decreases, and stabilizes at a plateau after approximately 200 iterations.

The convergence history of the volume fraction constraint in Figure 7(b) appears more fluctuating than the elastic compliance in Figure 7(a). A plausible explanation is that when the constraint is satisfied (or violated) in one iteration, it leads to a correspondingly small (or large) penalty term in the Lagrangian formulation [32]. This, in turn, can cause the constraint to be violated (or satisfied) in the following iteration, resulting in oscillatory behavior. To better manage design constraints, future work will incorporate more advanced constrained optimization techniques, such as the augmented Lagrangian method [59-61], sequential quadratic programming [45], method of moving asymptotes [62] or Newton-based methods [45].

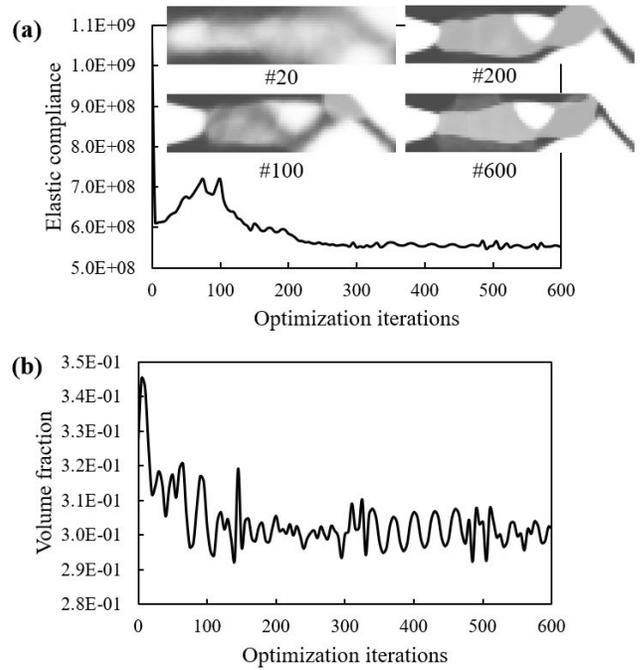

**Figure 7 Convergence histories of the cantilever beam experiment: (a)** The optimization history of elastic compliance along with four demonstrated topologies at selected iterations; **(b)** Convergence history of the volume fraction constraint.

To demonstrate material spatial distribution, we assemble the optimized structure from Figure 6(d) with its associated composite microstructures that are reconstructed from the design parameters, as shown in Figure 8(a). We include three snapshots of microstructures at difference locations as in Figure 8(b)-(d). On the one hand, high concentrations of the stiffer constituent (color coded as blue) are optimized to allocate at the top and bottom edge regions that are subject to the high magnitudes of bending-induced tension and compression forces. On the other hand, high volume fractions of the softer material (color coded as green) are found in the middle region. Both regions shown in Figure 8(b) and Figure 8(c) manifest isotropic material distributions with smooth transitions across microstructure boundaries. Additionally, the interior region of the truss-like section close to the force boundary condition shows high volume fractions of stiffer constituent which is gradually replaced by the softer constituent towards outer surfaces. We can see the distinctions of morphologies and volume fractions among the $3 \times 3$ microstructures in Figure 8(d) where smooth material transitions are enforced across microstructure boundaries.

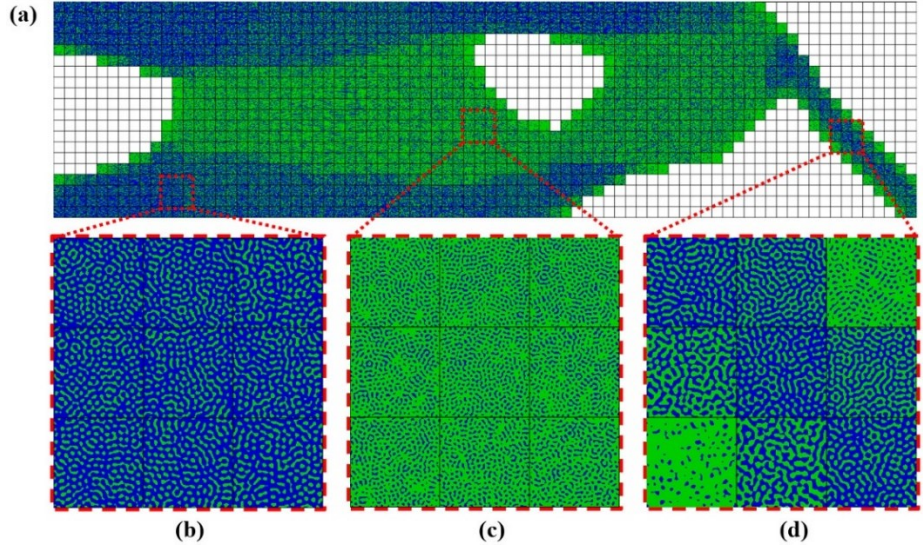

**(a)**

**(b)**   **(c)**   **(d)**

**Figure 8 Optimized topology and material distributions of the cantilever beam:** Spatially varying composite microstructures are assembled on the optimized structure topology where high volume fractions of the stiffer composite constituent (color coded as blue) are found along the top and bottom edges of the cantilever beam with large nominal (tension and compression) stresses, while the softer material (color coded as green) is allocated in the middle region.

A key advantage of our TO framework is its capacity for transfer learning, the ability to reuse the trained NNs' weight parameters from one TO experiment on the same model but with a different mesh resolution. To demonstrate this, we extract the NN weight parameters from the TO computation on an $80 \times 20$ mesh in Figure 6(d), and apply them to three alternative meshes: a $20 \times 5$ mesh in Figure 9(a), a $40 \times 10$ mesh in Figure 9(b), and a $160 \times 40$ mesh in Figure 9(d).

When comparing the resulting topologies across these meshes, we observe a high degree of similarity, indicating that the NNs have effectively learned a mapping from spatial coordinates to optimal topologies. However, examining the material distributions at corresponding locations within the design domain reveals a distinct trend: as mesh resolution increases, the material morphology transitions from relatively homogeneous (Figure 9(a)) to increasingly complex and spatially varied patterns in Figure 9(b)-(d). This occurs because the NNs were trained on a medium-resolution mesh; projecting this solution onto lower-resolution meshes results in information loss and simplified designs. Conversely, applying the learned weights to a finer mesh enhances detail, producing a bowl-shaped material distribution with smooth transitions across microstructural boundaries. Understanding the theoretical relationship between coarse and fine mesh designs during this transfer learning remains an open question and will be explored in our future work.

It is important to note that the transfer learning process involves only a single forward pass, where the trained NNs map the element centroid coordinates to their corresponding design parameters, $\rho_M$, $\rho_m$, $R_{out}$, and $\Delta R$. These parameters are then used to reconstruct the appropriate microstructures, which are assembled to form the optimized structure. This process is highly efficient, as it bypasses computationally intensive steps such as GP regression, nonlinear FEA, and constrained optimization. As a result, the method offers a fast and flexible solution for adapting designs to varying mesh resolutions.

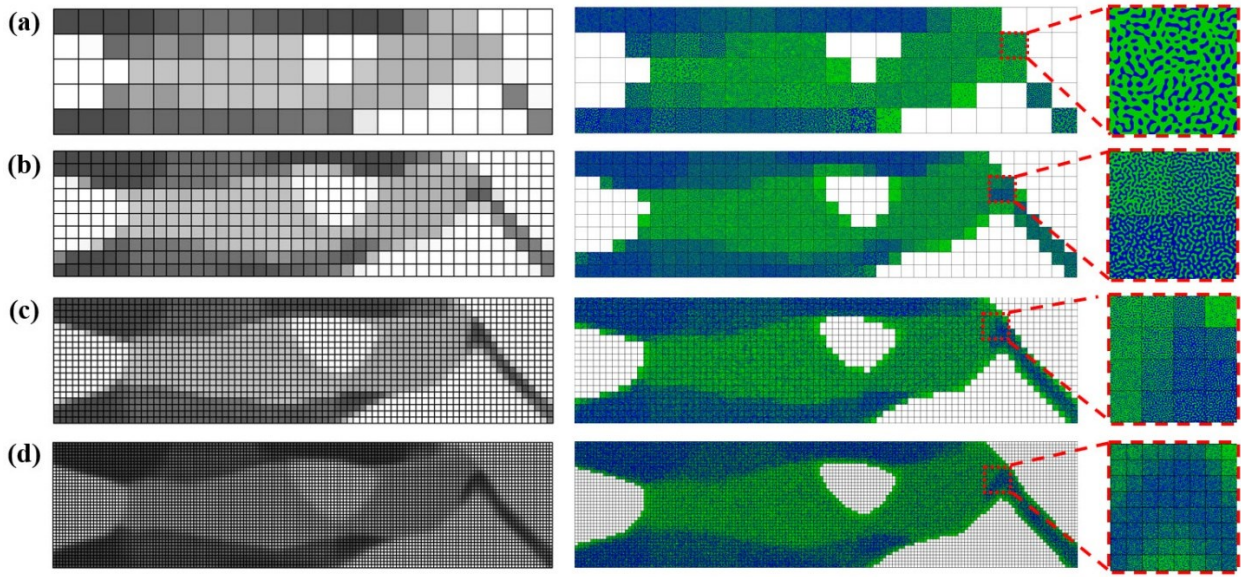

**Figure 9 Demonstration of neural TOPO's transfer learning capacity:** The NNs trained from the $80 \times 20$ macro-mesh in Figure 8 are applied to the same cantilever beam but with different mesh densities: **(a)** Inference results on a $20 \times 5$ mesh includes the optimized structural topology and material distributions; **(b)** Inference results on a $40 \times 10$ mesh; **(c)** The original solutions from the $80 \times 20$ mesh; **(d)** Inference solutions on a $160 \times 40$ mesh. Enlarged microstructures are demonstrated for the same location on the design domain.

## 7.2. Double-Clamped Beam

In this experiment, we continue to test our multiscale TOPO on a double-clamped beam with a non-zero Dirichlet boundary condition and we aim to maximize its total potential energy as design objective.

As shown in Figure 10(a), we fix the beam at both left and right edges and apply a displacement boundary condition $D$ at the center of its top edge. The dimension of the beam is $60 \times 20$, and it is discretized by 1200 quadrilateral finite elements of unit length.

To demonstrate the importance of nonlinear TOPO, we perform a linear elastic TOPO and let it be subject to a large deformation by setting $D = 20$. We show the optimized topology and its deformed configuration in Figure 10(b) and Figure 10(c), respectively. Although the triangle-shaped design closely resembles its small-deformation counterpart, due to that the linear elastic FEA relying solely on the initial geometry, it undergoes significant buckling under large deformation. As a result, it fails to provide sufficient structural support for continued loading. This example demonstrates that linear elastic designs are inadequate for applications involving large deformations and highlights the necessity of adopting nonlinear TO methods in such scenarios.

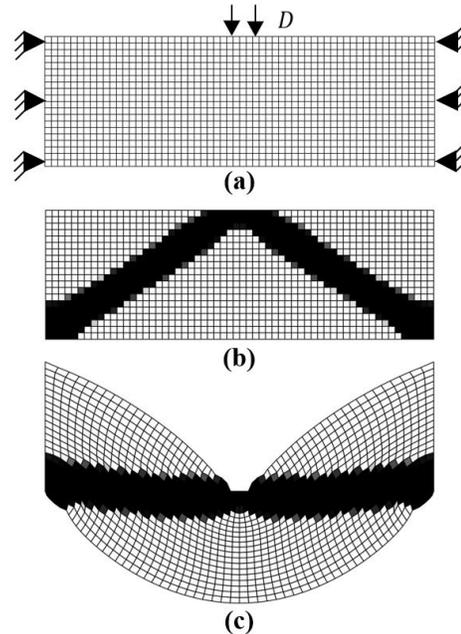

**Figure 10 Design setup of the double clamped beam experiment: (a)** The design domain is discretized by a $60 \times 20$ mesh with quadrilateral elements of unit length. The domain is fixed at both left and right edges and subject to a displacement boundary condition ($D$) at the center of the top edge; **(b)** The single-scale linear elastic TOPO design with $D = 20$; **(c)** Its deformed (buckled) configuration.

For comparison, we perform the same optimization experiment but using the single-scale single-material hyperelastic TOPO model Figure 11. Specifically, we increase the magnitudes of displacement boundary conditions from a small deformation of $D = 0.1$ in Figure 11(a) to large deformations of $D = 5$, $D = 10$ and $D = 20$ in Figure 11(b)-(d), respectively. Compared to the linear elastic design in Figure 10(b)-(c), we observe that the optimal topologies are significantly different for different magnitudes of boundary conditions. It is due to the nonlinear FEA discussed in Section 2 that takes into the account of the deformed configurations.

When we compare the four topologies in Figure 11, we can see that the triangle-shaped design in Figure 11(a) is similar to the linear elastic design in Figure 10(b), as it can effectively resist small deformation. When we increase the displacement boundary condition to $D = 5$ in Figure 11(b), the W-shaped design appears more effective than the triangle-shaped design in resisting buckling under large deformations. When the deformation is increased to $D = 10$ in Figure 11(c), the cavity in the middle of the structure is filled with material to resist the large deformation.

As we continue to increase the displacement boundary condition to $D = 20$ as Figure 11(d), we observe that the material in the bottom section of the W-shaped design is relocated to the center region to further improve the structural resistance to large deformation. By studying the deformed configurations of the optimized topologies under different loading conditions, we find that the nonlinear TOPO designs are more effective than their linear elastic counterparts to withstanding large displacements and extreme distortions while avoiding buckling.

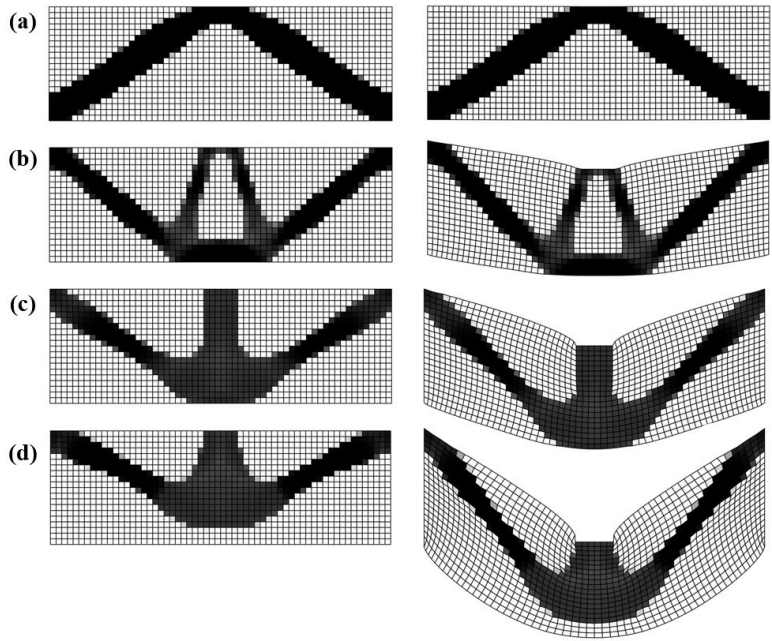

**Figure 11 Single-scale hyperelastic TOPO of the double clamped beam: (a)** The undeformed and deformed configurations of the optimized design under the displacement boundary condition of $D = 0.1$; **(b)** The design under $D = 5$; **(c)** The design under $D = 10$; **(d)** The design under $D = 20$.

We apply our multiscale TOPO to design the double-clamped beam under different displacement boundary conditions as shown in Figure 12. On the one hand, for the small displacement boundary condition ($D = 0.1$), the multiscale design of Figure 12(a) manifests a similar triangle-shaped topology as its single-scale counterpart of Figure 11(a). Additionally, it also has two truss-like regions attached to the top left and right corners. As we increase the deformation to $D = 5$, the optimized structure in Figure 12(b) shows a similar W-shaped topology as its single-scale counterparts in Figure 11(b). In addition, it has two truss-like regions attached to the bottom left and right corners. The plausible reason for the appearance of the additional truss-like regions in Figure 12(a)-(b) is that our composite microstructures have distinct material properties from the single type of material used in the single-scale design. On the other hand, when we continue to increase the displacement boundary condition to $D = 10$ and $D = 20$ as shown in Figure 12(c) and Figure 12(d), we obtain similar W-shaped designs as their single-scale counterparts. Compared to the design of $D = 10$, the optimized structure of $D = 20$ relocate its materials from the bottom center region to fill the cavity in the center region to withstand larger deformations.

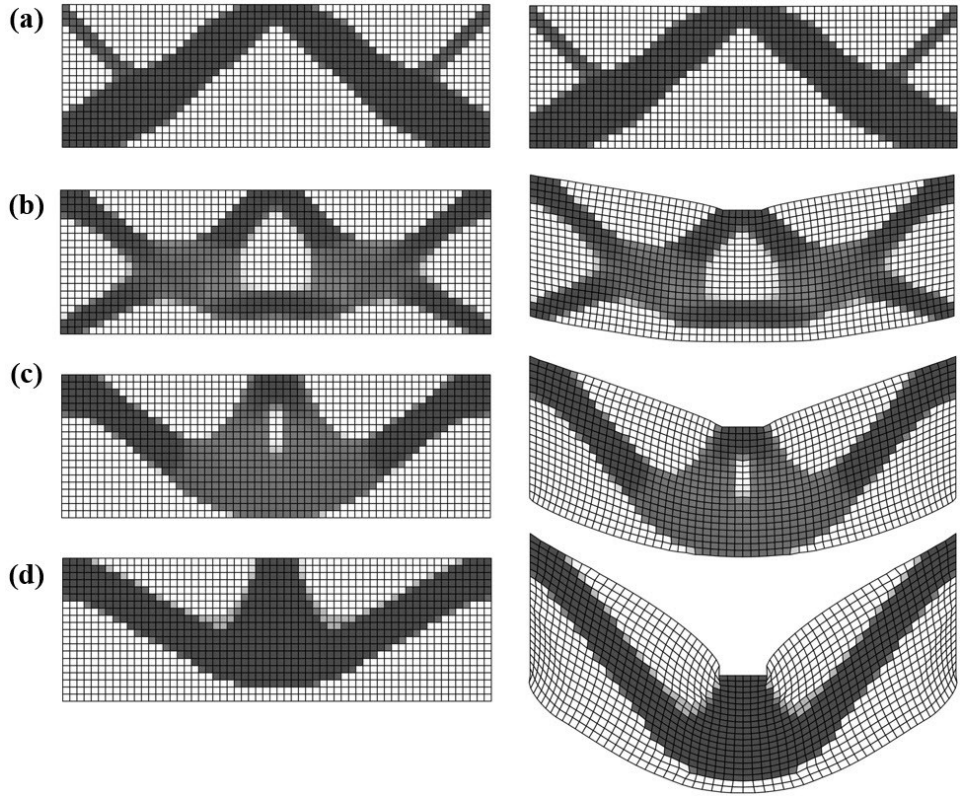

**Figure 12 Multiscale hyperelastic TOPO of the double clamped beam where each macro-element is assumed to be associated with a composite microstructure: (a)** The optimized structural topology and its deformed configuration under the displacement boundary condition of $D = 0.1$; **(b)** The design under $D = 5$; **(c)** The design under $D = 10$; **(d)** The design under $D = 20$.

To better understand our nonlinear TOPO model, we plot three field solutions as shown in Figure 13. Specifically, the interpolation coefficient of constitutive models is shown in Figure 13(a). The elements of zero value (color coded as blue) are governed by the linear elastic constitutive model, while the behaviors of the elements with unit coefficient value (color coded as yellow) are governed by the hyperelastic material model. By comparing to the deformed configuration in Figure 12(c), we find that the elements of non-zero volume fractions tend to have non-zero interpolation coefficients as long as they are not subject to extreme deformations, as described by Equation (47). We also demonstrate the Green-Lagrangian strain and PK2 stress in the Figure 12(b) and Figure 12(c), respectively. We observe the magnitude of strain can be up to 35% around the top left and right corners, and large stresses are found not only at the strain concentration regions but also at the center of the top edge where the non-zero displacement boundary condition is applied. The magnitude of PK2 stress at each macro-element is influenced by the effective material properties of its corresponding microstructure.

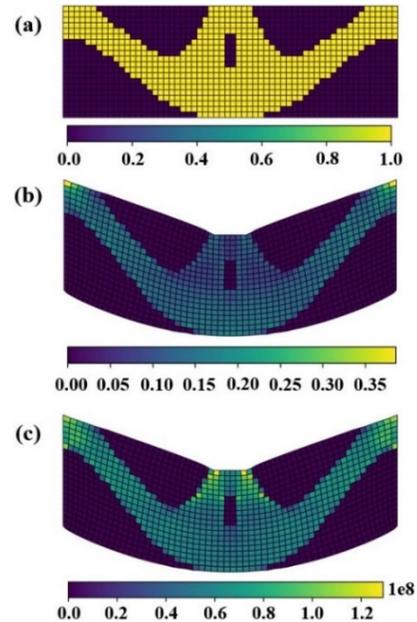

**Figure 13 Field solutions of the optimized double clamped beam with the boundary condition of $D = 10$:** (a) The interpolation coefficient between linear elastic and hyper elastic constitutive models in the undeformed configuration; **(b)** The Green strain in the deformed configuration; **(c)** The PK2 stress in the deformed configuration.

We reconstruct and assemble microstructures from the optimized design parameters and demonstrate the spatially varying material distributions as in Figure 14. High volume fractions of the stiffer constituent (color coded as blue) are found to distribute at the strain and stress concentration regions to withstand large deformation-induced internal forces, as in Figure 14(a). We also find that the usage of the stiffer constituent is gradually reduced in less-stressed areas, e.g., at the bottom of the center cavity in Figure 14(b) and towards outer surfaces in Figure 14(c). At all three enlarged snapshots in Figure 14, we see not only the spatially varying material morphology but also the smooth interface transition across microstructure boundaries.

We compare the total potential energies of the single-scale and multiscale TOPOs at different displacement boundary conditions in Figure 15. Both single scale and multiscale designs manifest clear hyperelastic responses. Additionally, the multiscale designs can achieve higher total potential energy values, therefore better than their single scale counterparts as in Equation (52).

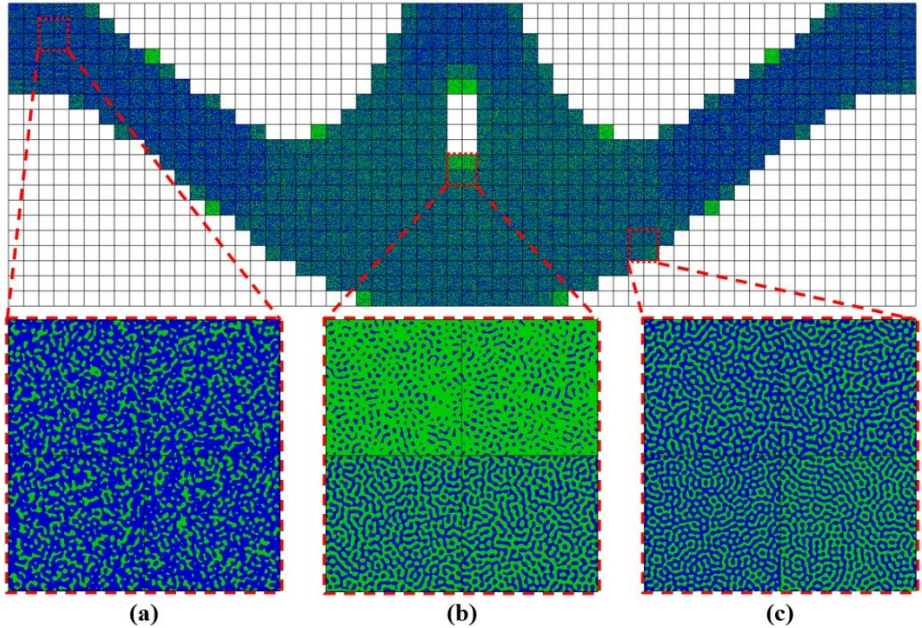

**(a)**             **(b)**             **(c)**

**Figure 14 Optimized topology and material distributions of the double clamped beam:** Spatially varying composite microstructures are assembled on the component. **(a)** High volume fractions of the stiffer composite constituent (color coded as blue) are allocated on the arm-like regions close to the left and right edges and around the center of the top edge, where non-zero displacement boundary conditions are applied. Spatially varying volume fractions and morphologies are found in regions with low stress concentrations, such as **(b)** and **(c)**.

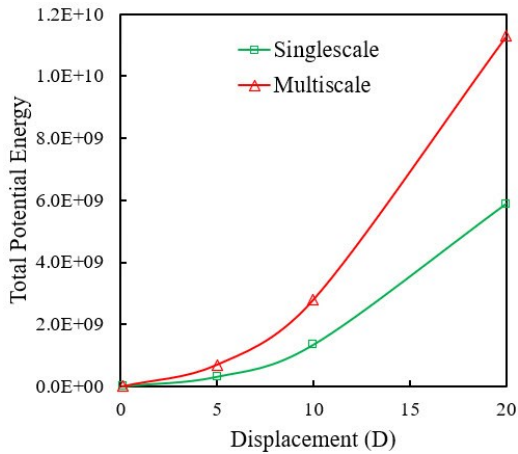

**Figure 15 Comparison of total potential energies between the single-scale and multiscale hyperelastic TOPOs**: Multiscale designs demonstrate higher total potential energies than their single-scale counterparts when subject to the same displacement boundary conditions.

### 7.3. Square Beam

The design region of the square beam is shown in Figure 16(a). It is discretized by a $40 \times 40$ finite element mesh where each element has a unit length. While the square beam is fixed at left edge, a non-zero displacement boundary condition $D$ is applied at the center of right edge.

We first demonstrate the designs of the single scale TOPOs when we set the displacement boundary condition as $D = 0.1$ and assume materials follow linear elastic and hyperelastic constitutive models as in Figure 16(b) and Figure 16(c), respectively. We observe that under the small deformation, the hyperelastic design is almost identical to the linear elastic design, while the two designs achieve symmetry about the horizontal plane. The two topologies are similar because the Neo Hookean hyperelastic constitutive model can be reduced to the isotropic linear elastic Hooke's law under small deformations.

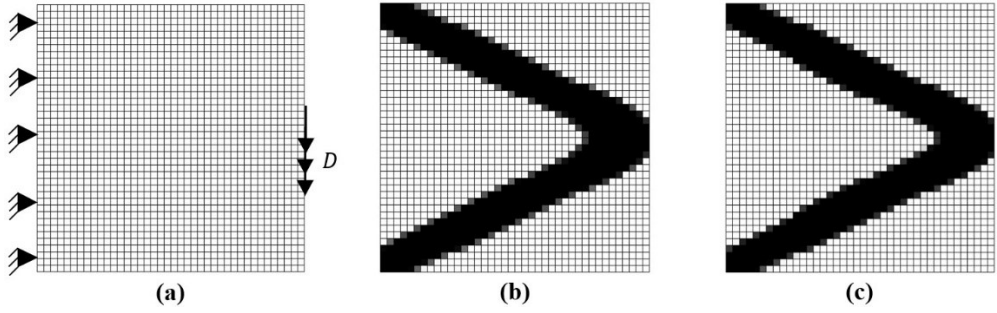

**Figure 16 Design setup of short cantilever beam experiment: (a)** The design domain is discretized by $40 \times 40$ quadrilateral elements with unit size length. The domain is fixed at the left edge, and a displacement boundary condition ($D$) is applied onto the center of the right edge. **(b)** The design of a single-scale linear elastic TOPO for $D = 0.1$. **(c)** The design of a single-scale hyperelastic TOPO for $D = 0.1$.

We increase the magnitudes of the displacement boundary conditions and study their influence on the designs. Specifically, we perform four single-scale hyperelastic TOPOs by increasing the magnitude of the displacement boundary condition $D$ from 1 to 10, 20, and 30, respectively, as shown in Figure 17. It is seen that as we gradually increase deformation magnitudes, the optimized topologies lose symmetry about the horizontal plane. More materials appear to allocate to the upper arm region to increase its resistance to the downward force. At the same time, the position of the lower arm region gradually moves towards the left side, from connecting to the center of the right edge as in Figure 17(a)-(b) to connecting to the middle of the upper arm as in Figure 17(c)-(d). Additionally, as we increase the boundary condition to $D = 20$, we observe that materials are removed from the bottom end of the lower arm and the lower arm moves upwards to avoid buckling under such large deformation, which is a similar phenomenon we observed from the previous experiment in Figure 11(d) and Figure 12(d).

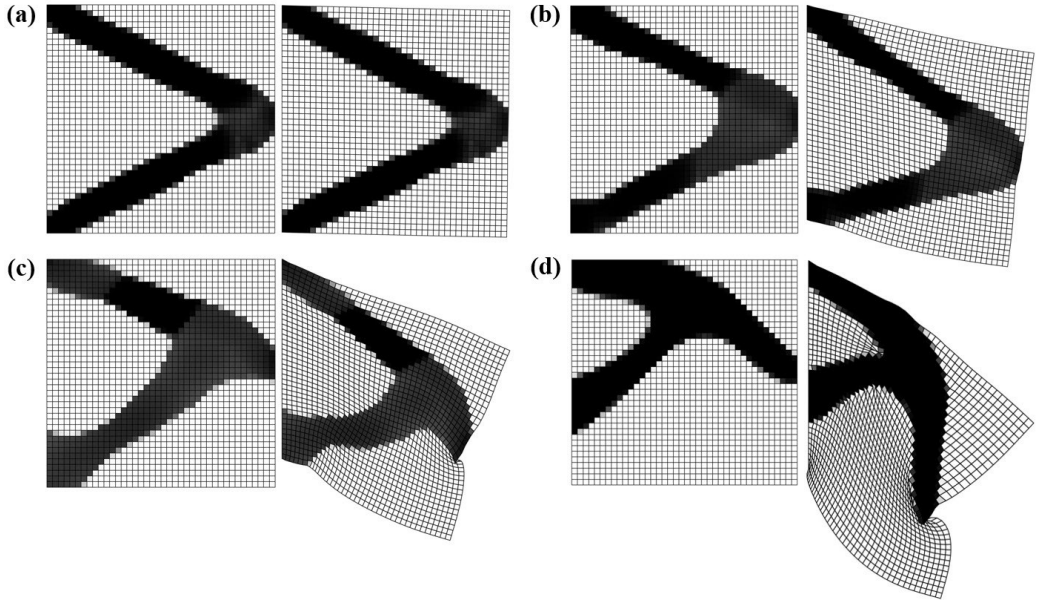

**Figure 17 Single-scale hyperelastic TOPO of the short cantilever beam: (a)** The undeformed and deformed configurations of the optimized design under the displacement boundary condition of $D = 1$; **(b)** The design of $D = 10$; **(c)** The design of $D = 20$; **(d)** The design of $D = 30$.

We perform the multiscale TOPOs with the same set of boundary conditions as shown in Figure 18. On the one hand, similar to the single scale designs in Figure 17, as we increase the magnitude of the displacement boundary conditions, the optimized structures lose their symmetries about the horizontal axes while allocating more materials to the upper arm section. On the other hand, we find that the position of the lower arm gradually moves to the position that connects to the mid-section of the upper arm as the displacement boundary condition is increased to $D = 20$. A difference is that the bottom end of the lower arm does not move upwards as in Figure 17(d). The reason is that the multiscale design has additional softer constituent, compared to the single-scale designs that can only utilize the stiffer composite constituent, to provide additional support and can effectively avoid buckling or extreme distortions on the lower arm region.

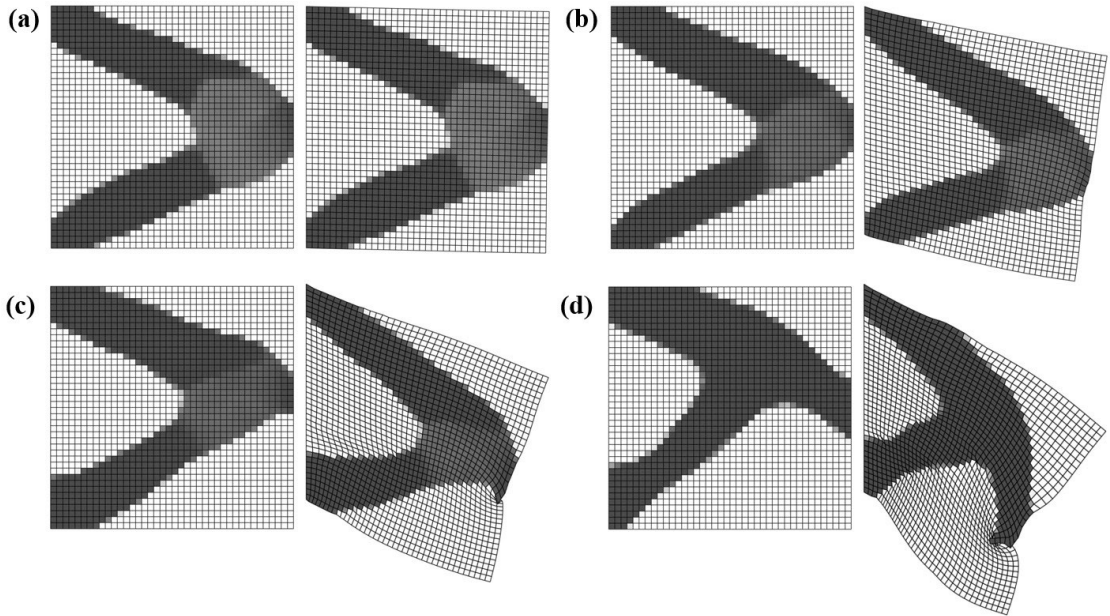

**Figure 18 Multiscale hyperelastic TOPO of the short cantilever beam where each macro-element is assumed to be associated with a composite microstructure: (a)** The optimized structural topology and its deformed configuration under the displacement boundary condition of $D = 1$; **(b)** The design under $D = 10$; **(c)** The design under $D = 20$; **(d)** The design under $D = 30$.

We plot the optimized multiscale design in Figure 19(d) corresponding to the large displacement boundary condition of $D = 20$ in Figure 18(d). We observe that the interior region of the structure is filled by microstructures with high volume fractions of the stiffer constituent. Its surface regions contain a layer of microstructures with softer constituent (color coded as green). The reason for such (mostly) uniform material distribution is that in the deformed configuration, the upper arm section rotates towards a vertical position such that its deformation is like an elongated tensile bar. For the lower arm section, it is mainly subject to the compression force passed from the upper arm. Its deformation is similar to a compression bar. Therefore, the microstructures in the upper and lower arm sections tend to have uniform morphology and volume fraction distributions, and the microstructures are optimized to select high volume fractions of the stiffer constituent to withstand the large deformation.

We create four histograms in Figure 20 to demonstrate the distributions of the four design parameters for all the 1600 microstructures, i.e., $\rho_M$, $\rho_m$, $R_{out}$ and $\Delta R$, for the multiscale design in Figure 18(d). First, the values of the macroscale volume fraction $\rho_M$ are converged to either zero or one as in Figure 18(a) due to the SIMP penalization scheme as in Equation (29). Second, nearly 1500 out of 1600 microstructures have the highest volume fraction ($\rho_m = 70\%$) of the stiffer composite constituent as in Figure 18(b). This is consistent with our observation from Figure 19 that the stiffer constituent is selected to dominate the microstructure to withstand tension and compression deformations in the upper and lower arm regions, respectively. Additionally, more than half of the microstructures (about 900 out of 1600) have the same value of $R_{out}$ of 25 as in Figure 20(c), and 450 out of 900 microstructures have the same value of $\Delta R$ as 25 in Figure 20(d). The distributions of the $R_{out}$ and $\Delta R$ together with $\rho_m$ are consistent with the multiscale design in Figure 19 that most microstructures have similar volume fractions and microscale material feature sizes in this design scenario.

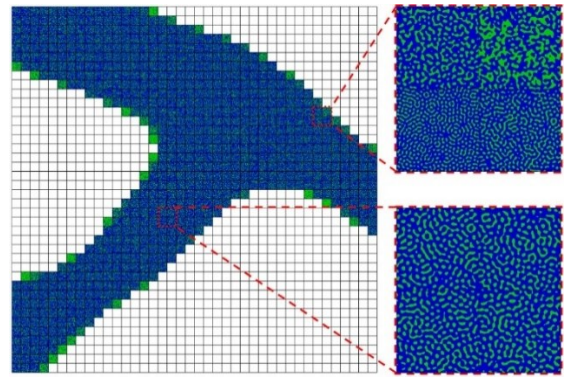

**Figure 19 Optimized topology and material distributions of the short cantilever beam for $D = 30$:** The stiffer constituent (color coded as blue) is distributed in the interior region of the design domain, while the softer material (color coded as green) is optimized to allocate close to the surface regions of the domain.

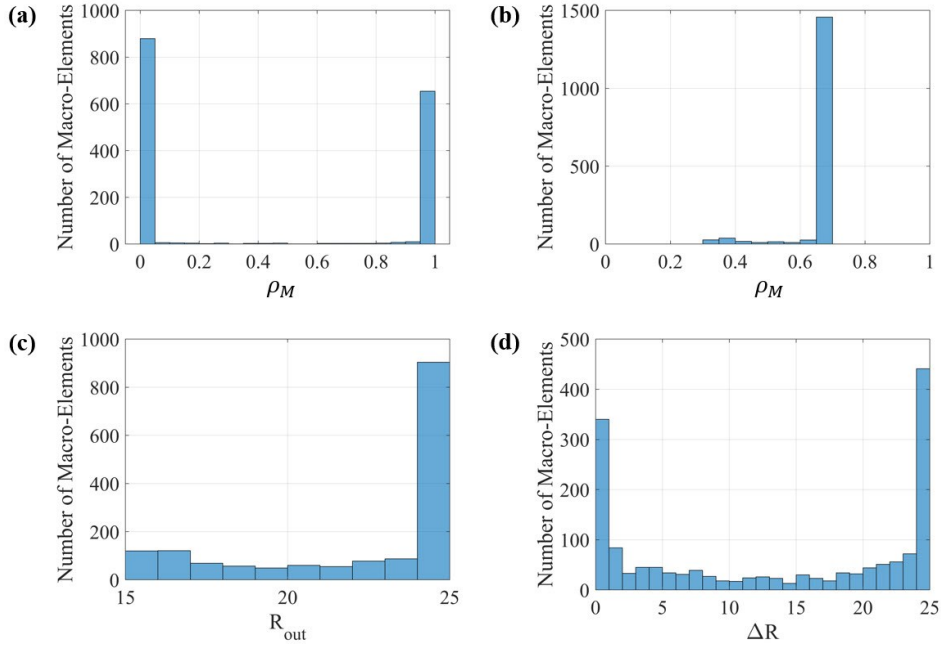

**Figure 20 Distributions of the design parameters for the 1600 macro-elements in the short cantilever beam experiment: (a)** The histogram of the macroscale volume fraction $\rho_M$; **(b)** The microscale volume fraction $\rho_m$; **(c)** SDF outer radius $R_{out}$; **(d)** The SDF radius intervals $\Delta R$.

## 8. Conclusion

We propose a general TO framework for designing multiscale FGMs composed of graded soft unit cell architectures that are capable of withstanding large deformations. Central to this approach is a data-driven surrogate model based on GP regression, which estimates the effective elastic properties of microscale composites as a function of low-dimensional design parameters. These effective properties are computed through computational homogenization, supported by a novel material homogenization scheme that expresses the strain energy function of the composite microstructure in terms of that of soft constituents. The surrogate model is then coupled with a NNs-based TO, allowing the multiscale design problem to be reformulated as an implicit function of the network's weight parameters. By leveraging adjoint method, implicit function theorem, and automatic differentiation in a differentiable programming paradigm, our approach enables automatic and efficient computation of nonlinear sensitivities for different design objectives and constraints.

We validate the proposed TO framework through a series of benchmark problems involving large deformations, including a cantilever beam (Section 7.1), a double-clamped beam (Section 7.2), and a square beam (Section 7.3). Our results highlight key distinctions between linear elastic and hyperelastic TOs, as well as between single-scale and multiscale, and single-material and FGMs designs. For small deformations, TOs based on linear elasticity and hyper-elasticity produce similar structural layouts. However, under large deformations, the linear elastic designs become unreliable, as they are based on initial configurations and often lead to unstable or buckling-prone designs. In contrast, our multiscale FGMs design yields distinct topologies and spatially varying material distributions that adapt more effectively to large deformations. Our designs also exhibit smooth transitions in material morphology across microstructural interfaces. Despite the high computational costs of nonlinear TOs, our framework supports efficient projection of existing solutions onto different mesh resolutions, provided the design domain and loading conditions remain unchanged. It allows for flexible adjustment of spatial resolutions to meet varying design requirements with little additional cost.

Our work is currently implemented and demonstrated for 2D cases; however, its extension to 3D scenarios is straightforward in the future. This work can be extended by incorporating other nonlinear elastic constitutive models, particularly anisotropic formulations such as hyperelastic anisotropic models [63], to enable the design of soft composite structures with locally oriented directions. Other research directions of interest include nonlinear TOs for multiphysics [64,65], compliant mechanisms [33], plasticity [41], damage [42,43,66], multiple objectives and constraints [59-61].

### Acknowledgments

Shiguang Deng would like to thank the support of the Faculty Startup Grant and the New Faculty Research Development Grant from the University of Kansas. The Northwestern team would like to acknowledge the acknowledge the support from the DARPA METALS program project RADICAL (HR0011-24-2-0302) and the NSF BRITE fellow grant (2227641). The authors would also like to thank the constructive suggestions from anonymous reviewers.

## Appendix A. Microstructure Reconstruction

The objective for our multiscale TO is to generate FGMs with spatially varying composite microstructures on the optimized structural layout. This is achieved by systematically varying microstructural design parameters to produce a diverse set of material morphologies. In this work, our microstructure reconstruction is performed using the Spectral Density Function (SDF) [44]. SDF characterizes how Fourier components are distributed across the frequency range, with higher frequencies capturing smaller features and lower frequencies capturing larger features from spatial domains. It provides a physically meaningful frequency-based low-dimensional representation of quasi-random materials [67-69].

Compared to the traditional truss-, beam- and plate-based mechanical metamaterials, SDF-based stochastic representations often manifest spinodal morphologies with unique features of stochasticity, aperiodicity, anisotropy, and irregularity. Such stochastic microstructures often exhibit superior properties to the conventional metamaterial designs due to their high resistance to stress concentrations at stress-raisers and material interfaces, and insensitivity to manufacturing imperfections and defects.

We note that there are multiple approaches to use SDF functions to construct material composition, morphology, and composite architecture, depending on processing techniques. While it is straightforward to create bulk materials with the same microstructure, it is difficult to produce structures with spatially varying microstructures. In this work, we use SDF representation to illustrate the use of data-driven approach for the design of FGMs with spatially varying microstructures without the consideration of manufacturing contexts.

Our SDF-based microstructure reconstruction is based on the Linear Time Invariant (LTI) model [70], which states that the output signal can be deduced from an impulse and the arbitrary input signal [71] as:

$$S_o(\xi) = |H|^2 S_i(\xi), \tag{A.1}$$

where $S_i$ and $S_o$ are the SDFs of input and output signals with the frequency of $\xi$, and H is the Fourier Transform of an impulse response. SDF is the squared magnitude of the Fourier Transformation of the signal $\varphi$ [71] as:

$$S(\xi) = |F[\varphi]|^2, \tag{A.2}$$

where $F[\cdot]$ is the Fourier Transform operator. Since the phase field of a microstructure can be considered as a 2D signal $\varphi$, the microstructure can be reconstructed from Equation (A.1), as follows:

$$S_R(\mathbf{k}) = S_T(\mathbf{k}) \bullet S_W(\mathbf{k}), \tag{A.3}$$

where $S_R$, $S_T$, and $S_W$ are the SDFs of reconstructed, target, and white noise signals (signals can be considered as the phase fields of microstructures), and $\mathbf{k}$ is the spatial frequency vector. That is, $\mathbf{k}$ is the vector version of the scalar frequency of $\xi$ in Equation (A.1). The operator " $\bullet$ " represents point-wise multiplication, and the reconstructed, target, and white noise microstructures must have the same spatial resolutions. By comparing Equation (A.1) with Equation (A.3), it is clear the target phase field $S_T$ resembles the impulse $H$, while the reconstructed and white noise phase fields ($S_R$ and $S_W$) correspond to the input and output signals ($S_i$ and $S_o$), respectively. The phase field of the reconstructed microstructure can be computed as:

$$\varphi_R = \left| F^{-1}\left[ \sqrt{S_T(\mathbf{k})} \bullet F[\varphi_W] \right] \right|$$
$$= \left| F^{-1}\left[ |F[\varphi_T]| \bullet F[\varphi_W] \right] \right|, \tag{A.4}$$

where the phase field of the reconstructed microstructure $\varphi_R$ can be computed by applying Fourier transformations on the phase fields of the target microstructure $\varphi_T$ and white noise microstructure $\varphi_W$, respectively.

It is noted that white noise microstructure $\varphi_W$ introduces stochasticity, since there exist multiple different realizations of $\varphi_W$ depending on the probabilistic distribution of the white noise. Additionally, the white noise microstructure $\varphi_W$ can be considered as a white noise image containing all frequencies. The reconstruction process in Equation (A.4) can be therefore interpreted as follows: the target signal $\varphi_T$ performs an convolution operation on the white noise signal $\varphi_W$ to filter out all frequencies except for the target frequencies in $\varphi_T$. For this reason, the phase field of the reconstructed microstructure $\varphi_R$ is considered as a random realization of $\varphi_T$.

We use a level-set function to cut the phase field $\varphi_R$ to obtain the binary image of the reconstructed microstructure:

$$\Lambda(\mathbf{X}) = \begin{cases} 1, & \text{if } \varphi(\mathbf{X}) \leq \varphi_{cut}(\rho) \\ 0, & \text{otherwise,} \end{cases} \tag{A.5}$$

where $\Lambda(\mathbf{X})$ is the binary image of the reconstructed microstructure at spatial position $\mathbf{X}$ with only two values, i.e., either 1 when the phase field $\varphi$ is smaller or equal to the cutting plane $\varphi_{cut}$ (corresponding to the desired microstructural volume fraction $\rho_m$), or 0 when the $\varphi$ is larger than the cutting plane value. If $\varphi_T$ is periodic, the microstructure $\Lambda(\mathbf{X})$ preserves periodicity.

By Equations (A.4) and (A.5), we can reconstruct isotropic microstructures by three design parameters as shown in Figure 21(a): the volume fraction of a composite constituent $\rho_m$ (e.g., the constituent color coded as blue in Figure 21(a)), the outer radius of the SDF ring $R_{out}$, and the radius interval between outer and inner radii $\Delta R$ in Figure 21(b). We note that the dimensions of the microstructure and the SDF field are the same as described by Equation (A.3). In Figure 21(a)-(b) both are represented by $500 \times 500$ pixels. We also note that the spatial frequency $\mathbf{k}$ of Equation (A.3) is shown as red-colored pixels on the radius of the hollow ring on the SDF field. The isotropic 2D SDF field can be converted to a 1D function [44] as shown in Figure 21(c) where all the SDF values are enforced as zero (color coded as black) except for the target frequencies at the center (with zero frequency) and on the ring of the SDF field.

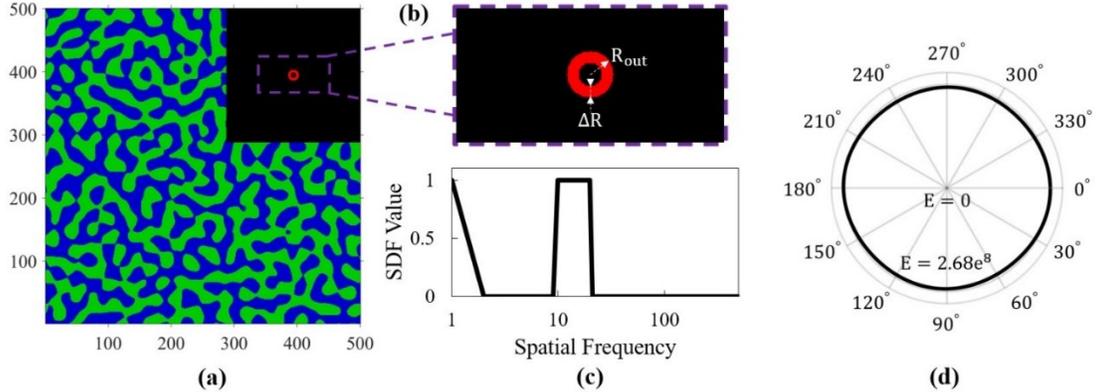

**Figure 21 Design of composite microstructures using SDF parameters: (a)** The composite microstructure consists of two hyperelastic constituents, i.e., a stiffer phase (color-coded as blue) and a softer phase (color-coded as green), along with the microstructure's corresponding SDF field; **(b)** The ring-shaped 2D SDF is parameterized by two design parameters: the outer radius ($R_{out}$) and the interval between outer and inner radii ($\Delta R$); **(c)** The 2D SDF can be converted to a 1D SDF function of spatial frequencies where only the frequencies on the ring and at the center have a unit SDF value and color coded as red; **(d)** The elastic surface of the effective Young's modulus of the microstructure along different directions.

We use computational homogenization [38] to calculate the elastic surface of the effective Young's modulus of the microstructure in Figure 21(d) (see Section 7 for detailed material properties). The elastic surface manifests an isotropic state of the elastic modulus. We therefore can apply the isotropic Neo Hookean model from Section 2 to simulate homogenized behaviors of reconstructed microstructures. We plot several reconstructed microstructures along with associated 2D SDF fields to demonstrate the effects of the three microstructure design parameters on the microstructure composition and morphology as shown in Figure 22.

In the first row of Figure 22, we illustrate the influence of the interval radius of the SDF, i.e., $\Delta R$. Specifically, on the microstructures with the dimension of $500 \times 500$, we set the values of the SDF outer radius and the volume fraction as $R_{out} = 20$ and $\rho_m = 0.5$, respectively, and change the value of $\Delta R$ to 20, 15, 10 and 1 as in Figure 22(a1)-(a4). We observe as $\Delta R$ decreases, the SDF gradually changes from a solid circle to hollow rings. That is, as the number of spatial frequencies (shown as the red pixels) reduces, reconstructed microstructures manifest more regular patterns (similar frequency magnitude) with smaller feature sizes (lower frequency magnitudes are removed).

We demonstrate the influence of the SDF outer radius $R_{out}$ in the second row of Figure 22. We set the values of the SDF radius interval and the material volume fraction as $\Delta R = 5$ and $\rho_m = 0.5$, while changing the value of $R_{out}$ to 10, 20, 30, and 40, respectively. We observe that as we increase $R_{out}$ (resulting in larger frequency magnitudes), material feature sizes in reconstructed microstructures gradually reduce, as shown from Figure 22(b1) to Figure 22(b4). It is well established that microstructures with sufficiently small features, specifically, those where the overall size is at least an order of magnitude larger than the feature scale, yield consistent and reliable effective properties through computational homogenization [72,73,41]. To ensure scale separation, we carefully select the range of the SDF outer radius $R_{out}$ when constructing the data repository of reconstructed microstructures in the next section.

To demonstrate the influence of the volume fraction, we fix the values of the outer and inner radii of the SDF fields by setting $R_{out} = 20$ and $\Delta R = 5$ but gradually change $\rho_m$ as 0.2, 0.4, 0.6 and 0.8, respectively, as in Figure 22(c1)-(c4). As we change the volume fraction, the microstructures show different morphologies. Specifically, when the volume fraction of one constituent significantly exceeds that of the other, as shown in Figure 22(c1) and Figure 22(c4), the minority phase tends to form isolated regions, resulting in scattering patterns characterized by dispersed islands. However, when the volume fractions of the two constituents are close, e.g., in Figure 22(c2) and Figure 23(c3), both materials tend to have bi-continuous morphologies.

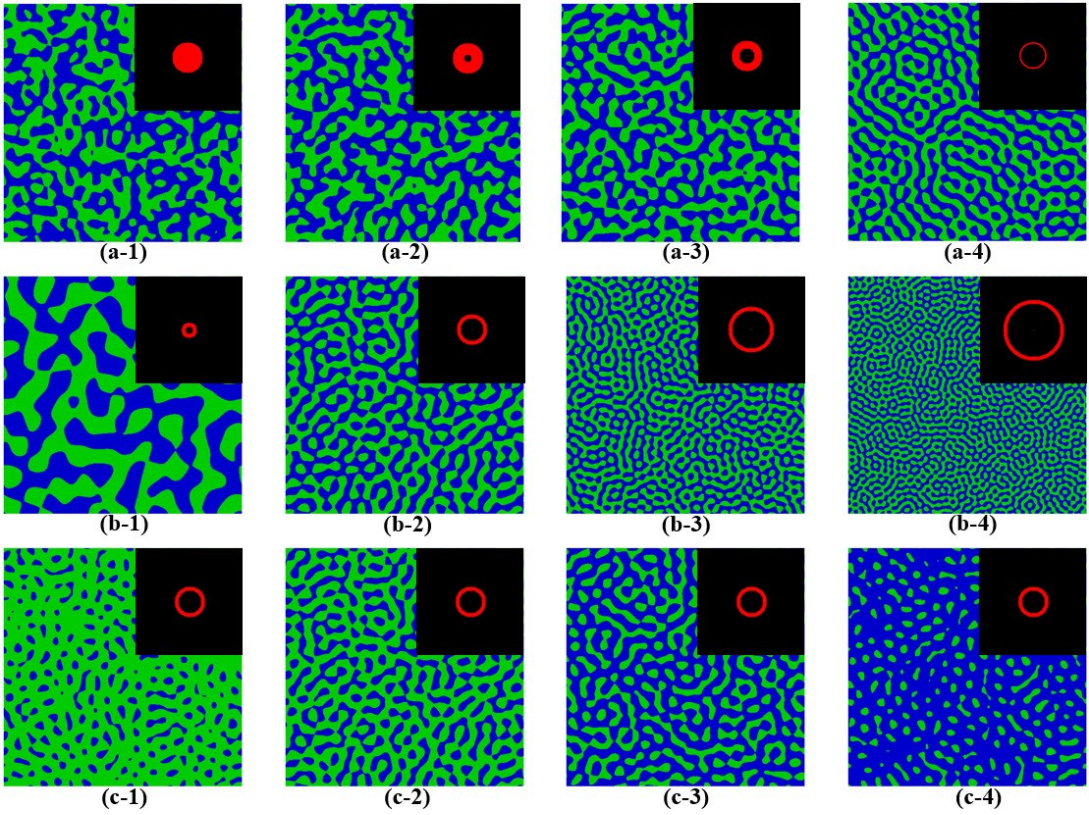

**Figure 22 Composite microstructures created by different values of design parameters: (a)** Four microstructures are generated using the same SDF outer radius ($R_{out} = 20$) and volume fraction ($\rho_m = 0.5$) but different values of SDF radius intervals of ($\Delta R = 20, 15, 10, 1$), respectively; **(b)** Four microstructures are generated using $\Delta R = 5$ and $\rho_m = 0.5$, but different outer radii of ($R_{out} = 10, 20, 30, 40$), respectively; **(c)** Four microstructures are generated using $R_{out} = 20$ and $\Delta R = 5$, but different volume fractions ($\rho_m = 0.2, 0.4, 0.6, 0.8$) of a composite constituent (color coded as blue), respectively. Note that all 2D SDF fields are enlarged by three times for visualization.

To demonstrate the impacts of SDF parameters on the properties of reconstructed properties, we compute the L2 norms of the elastic tensors of the reconstructed microstructures of Figure 22 as in Equation (A.6), and we compare their values in Table 2.

$$\|\boldsymbol{C}\|_2 = \sqrt{C_{11}^2 + C_{22}^2 + C_{33}^2 + +C_{12}^2 + C_{13}^2 + C_{23}^2} \ . \qquad (A.6)$$

From Table 2 and Figure 22, we observe that as we decrease the SDF interval radii $\Delta R$, the morphologies of the reconstructed microstructures in Figure 22(a-1)-(a4) gradually change to more regular patterns with smaller feature sizes, while the norms of their elastic tensors $\|\boldsymbol{C}\|_2$ manifest similar but different values. Additionally, as we increase the SDF outer radii $R_{out}$ in Figure 22(b-1)-(b4), we observe feature sizes of reconstructed microstructures gradually reduce, resulting in convergent effective elastic moduli of $\|\boldsymbol{C}\|_2$. In the last, when we increase the volume fraction of the stiffer constituent $\rho_m$ in Figure 22(c-1)-(c4), we can see a clear increase in $\|\boldsymbol{C}\|_2$ indicating the strong influence of $\rho_m$ on the effective elastic properties.

**Table 2** L2 norms of the elastic tensors of the reconstructed microstructures in Figure 22.

| Microstructures | a-1 | a-2 | a-3 | a-4 | b-1 | b-2 | b-3 | b-4 | c-1 | c-2 | c-3 | c-4 |
|---|---|---|---|---|---|---|---|---|---|---|---|---|
| $\|\boldsymbol{C}\|_2$ | 4.470e8 | 4.467e8 | 4.436e8 | 4.481e8 | 4.419e8 | 4.530e8 | 4.517e8 | 4.523e8 | 2.297e8 | 3.458e8 | 5.921e8 | 1.045e9 |

### Appendix B. Microstructure Boundary Transition

Due to the periodic phase fields as in Equation (A.4), our stochastic microstructures appear periodic as shown in Figure 21(a) and Figure 22. Such periodicity is beneficial for microstructure analysis but imposes challenges for multiscale TO with spatially varying microstructures. Different morphologies in microstructures cause an ill-connected interface across boundaries. Such discontinuity may result in manufacturing difficulty and stress singularity. An example of an ill-connected interface is demonstrated in Figure 23(a), where the two microstructures, despite being composed of similar materials, exhibit distinctly different morphologies, resulting in a pronounced material discontinuity across their boundary (highlighted by the red box).

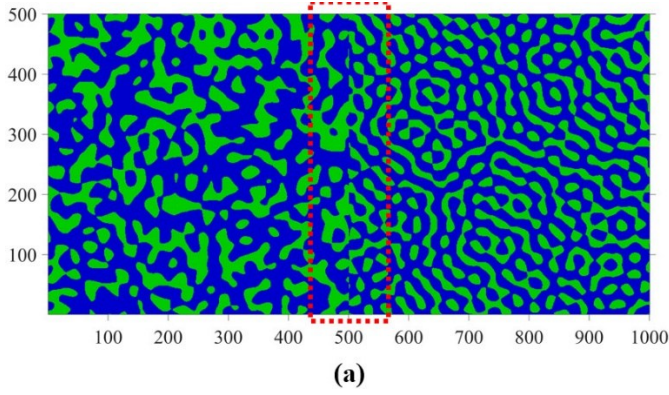
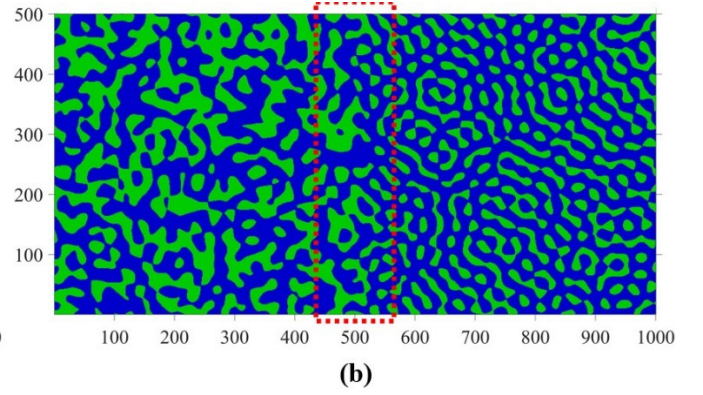

**(a)**           **(b)**

**Figure 23 Interface transition across microstructure boundaries: (a)** Material discontinuity is observed on the boundary at $x = 500$ of two periodic microstructures where each microstructure has a side length of 500 pixels. The boundary interface is highlighted inside a red box. **(b)** Well-connected boundary interface after applying interface interpolation and amplifying functions.

We develop a two-step approach to change the ill-connected interface in Figure 23 (a) to a smooth interface as shown in Figure 23(b).

In the first step, we apply an interface interpolation function to the phase fields of two neighboring microstructures as:

$$\varphi_{12}^{\text{int}}(\boldsymbol{X}) = \big(1 - \lambda(x)\big)\varphi_1(\boldsymbol{X}) + \lambda(x)\varphi_2(\boldsymbol{X}) \text{, and} \quad \text{(B.1)}$$

$$\lambda(x) = \frac{e^{-\zeta(x-2l)^2}}{e^{-\zeta(x-l)^2} + e^{-\zeta(x-2l)^2}} \text{,} \quad \text{(B.2)}$$

where $\varphi_{12}^{\text{int}}$ is the interpolated phase fields, while $\varphi_1$ and $\varphi_2$ are the phase fields of to-be-connected microstructures. $\lambda(x)$ is the interpolation function depending on the axis value $x$ (horizontal or vertical axes). While $l$ represents the size of the microstructure, e.g., $l$=500 in Figure 23, $\zeta$ is a user-defined constant that decides the size of interpolation regions. The smaller the value of $\zeta$, the larger the interpolation regions are used in the two microstructures. When $\zeta$ becomes too small, the interpolation regions in the microstructures would become significantly large and can change their interior morphologies, affecting microstructures' effective properties. We adopt $\zeta$=5e-5 as the default value for interpolation in this work. We demonstrate the interpolation functions for the two microstructures in Figure 23(b), as in Figure 24.

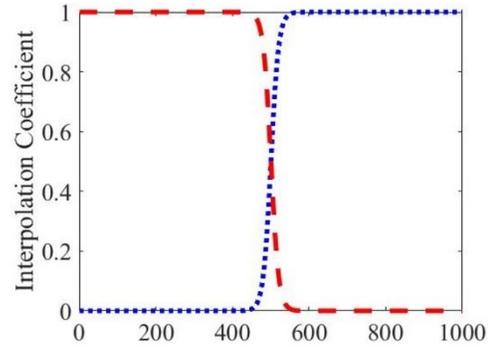

**Figure 24 Interface interpolation function:** Two interpolation functions are applied to the phase fields of the two microstructures of Figure 23(a) with $\zeta$=5e-5 in Equation (B.2).

In the second step, we introduce an interface amplifying function as follows:

$$\varphi_{12}^{\text{amp}}(\boldsymbol{X}) = \eta\big[1 + \gamma(x)\big]\varphi_{12}^{\text{int}}(\boldsymbol{X}) \text{, and} \quad \text{(B.3)}$$

$$\gamma(x) = \begin{cases} \lambda(x), & \text{if } x \leq l \\ 1 - \lambda(x), & \text{otherwise,} \end{cases} \quad \text{(B.4)}$$

where $\varphi_{12}^{\text{amp}}$ is the amplified phase field from the interpolation $\varphi_{12}^{\text{int}}$ in Equation (B.1), and $\gamma(x)$ is the amplifying function depending on the value $x$ along the interpolation axis. $\eta$ is the amplifying coefficient that controls the magnifying ratio for the phase fields on the boundary. A higher value of $\eta$ results in more well-connected interfaces but results in larger feature sizes on the interface. We choose $\eta = 1$ as the default value in this work, and demonstrate the interface amplifying functions of Equations (B.3)-(B.4) in Figure 25.

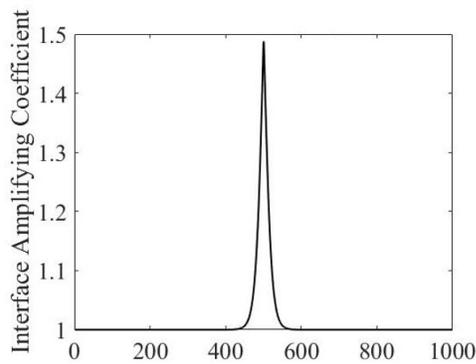

**Figure 25 Interface amplifying function:** Interface amplifying coefficients are applied along the interpolation direction. For the boundary interface at $x = 500$ in Figure 23(a), the phase fields of the two microstructures are multiplied by magnifying coefficients.